\documentclass[ aps,  twocolumn, nofootinbib, showkeys, notitlepage, superscriptaddress]{revtex4-2} 

\usepackage{natbib}
\usepackage{graphicx}
\usepackage{dcolumn}
\usepackage{bm}
\usepackage{amsmath}
\usepackage{amssymb}
\usepackage{float}
\usepackage{multirow}
\usepackage{slashed}
\usepackage{xcolor}
\usepackage{braket}
\usepackage{physics}
\usepackage{multirow}
\usepackage{gensymb}
\usepackage[colorlinks=true, pdfstartview=FitV, bookmarks=true, bookmarksnumbered=true, breaklinks]{hyperref}
\usepackage{mathtools,braket}

\usepackage{lipsum}  
\usepackage{color}
\definecolor{blue}{rgb}{0.0, 0.0, 1.0}
\definecolor{red}{rgb}{1.0, 0.0, 0.0}
\definecolor{royalblue}{rgb}{0.0, 0.14, 0.4}
\hypersetup{linkcolor=red, citecolor=blue, urlcolor=blue}

\def\orcid#1{\kern .08em\href{https://orcid.org/#1}{\includegraphics[keepaspectratio,width=0.7em]{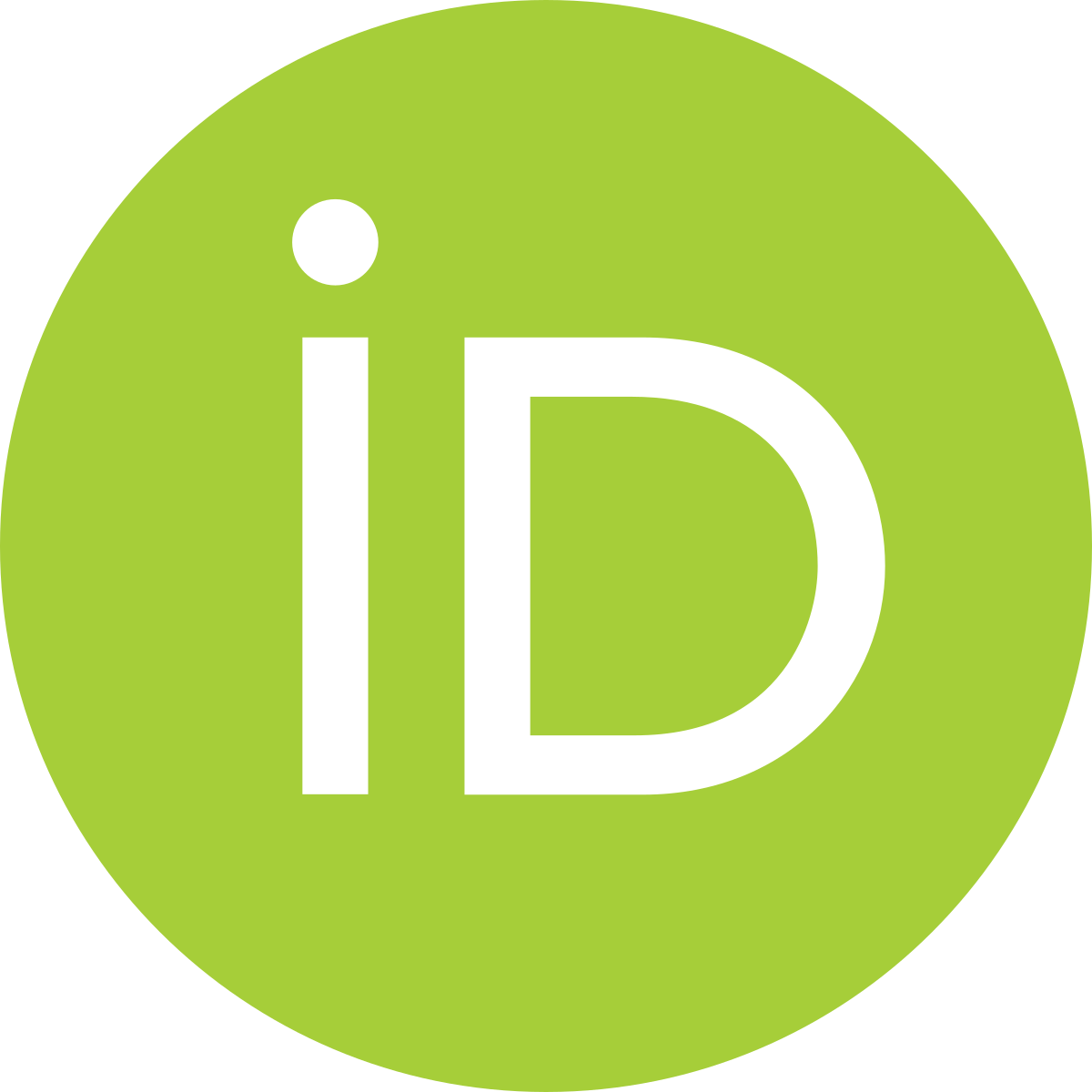}}}

\begin{document}
\title{Self-consistent $M1$ radiative transitions of excited $B_c$ and heavy quarkonia\\ with different polarizations in the light-front quark model}

\author{Muhammad Ridwan\orcid{0000-0002-2949-5866}}
\email{muhammad.ridwan75@sci.ui.ac.id}
\affiliation{Departemen Fisika, FMIPA, Universitas Indonesia, Depok 16424, Indonesia}

\author{Ahmad Jafar Arifi\orcid{0000-0002-9530-8993}}
\email{ahmad.arifi@riken.jp}
\affiliation{Few-Body Systems in Physics Laboratory, RIKEN Nishina Center, Wako 351-0198, Japan}
\affiliation{Research Center for Nuclear Physics (RCNP), Osaka University, Ibaraki 567-0047, Japan}

\author{Terry Mart\orcid{0000-0003-4628-2245}}
\email{terry.mart@sci.ui.ac.id}
\affiliation{Departemen Fisika, FMIPA, Universitas Indonesia, Depok 16424, Indonesia}

\date{\today}

\begin{abstract}

In this study, we investigate the properties of pseudoscalar and vector charmonia, bottomonia, and $B_c$ mesons using the light-front quark model, focusing on the $M1$ radiative transition. 
For that purpose, we conduct a variational analysis with a QCD-motivated effective Hamiltonian, employing a trial wave function expanded in the harmonic oscillator basis functions up to the $3S$ state. 
We fit the model parameters to mass spectra and decay constants, obtaining reasonable agreement with experimental data and correctly reflecting the hierarchy of mass spectra and decay constants. 
In analyzing the $M1$ radiative transition, we consider both good ($\mu=+$) and transverse ($\mu={\perp}$) current components with both longitudinal $(h=0)$ and transverse $(h=\pm1)$ polarizations, demonstrating that the results from both components of currents and polarizations are identical. 
Self-consistency is achieved by substituting $M$ with $M_0$ when computing the operators for decay constants and radiative transitions.
We also find that the difference between longitudinal and transverse polarizations of the observables may quantify the anisotropy of the model wave function.
Our results on radiative transitions align reasonably well with experimental data, lattice QCD, and theoretical predictions. Furthermore, we also provide predictions for $B_c$ mesons that can be tested in experiments.
\end{abstract}

\maketitle

\section{Introduction}

Understanding the fundamental components of the universe requires a deep knowledge of hadron physics. 
Governed by quantum chromodynamics (QCD), hadrons present significant challenges due to their nonperturbative nature, which complicates the study of their structure~\cite{Gross:2022hyw}.
Since the discovery of charmonium in 1974~\cite{E598:1974sol,SLAC-SP-017:1974ind}, these systems have been crucial for studying hadron structure due to their relatively simple composition, a pair of heavy quark and antiquark. 
In particular, investigating higher excited states of heavy quarkonia can provide insights into confinement.
For a comprehensive overview of the development of heavy quarkonia in various contexts, see Refs.~\cite{Brambilla:2010cs,Eichten:2007qx,Voloshin:2007dx,Patrignani:2012an,Chapon:2020heu,Lansberg:2019adr}.

To understand excited heavy quarkonia, it is essential to study various properties such as their mass spectra, decay constants, decay branching ratios, and transition form factors, as these properties provide unique and detailed insights into their internal structure. Radiative transitions, especially $M1$ transitions, provide a simple test of nonperturbative models and are valuable for probing the internal structure of quarkonia through their electromagnetic interactions.
The radiative transitions in quarkonia have been examined using the nonrelativistic potential model~\cite{Soni:2017wvy, Lakhina:2006vg, Brambilla:2005zw, Pineda:2013lta, Segovia:2016xqb, Deng:2016ktl,Deng:2016stx}, and some attempts have been made to include relativistic corrections~\cite{Li:2009zu, Zambetakis:1983te, Godfrey:1985xj, Grotch:1984gf, Godfrey:2001eb, Ebert:2002pp, Lahde:2002wj}. Studies on radiative transitions using various other methods have also been developed~\cite{Barnes:2005pb, Hong:2022sht, Ganbold:2021nvj, Issadykov:2023pwl, Simonis:2016pnh}. 
Moreover, lattice QCD~\cite{Dudek:2006ej, Dudek:2009kk, Hughes:2015dba, Delaney:2023fsc,Donald:2012ga,Becirevic:2012dc,Lewis:2012bh} has advanced the study of these transitions by providing providing tighter constraints on theoretical models.

Analyses of radiative transitions on the light front have been conducted using several approaches~\cite{Choi:2007se, Hwang:2006cua, Choi:2009ai, Ke:2010vn, Li:2018uif, Peng:2012tr, Shi:2016cef}; however, studies focusing on excited quarkonia remain limited~\cite{Ke:2010vn, Li:2018uif, Peng:2012tr, Shi:2016cef}. Notably, light-front dynamics(LFD)~\cite{Dirac:1949cp, Brodsky:1997de,Terentev:1976jk} provides an effective framework for addressing relativistic effects due to its unique rational energy-momentum dispersion relation. This approach incorporates the maximum number of kinematic (or interaction-independent) generators, simplifying the dynamics required to achieve QCD solutions that reflect full Poincaré symmetries.
Note that the approaches based on LFD has been applied to study various properties of mesons~\cite{Choi:1997iq,Choi:2015ywa,Arifi:2022pal,Pandya:2024qoj,Acharyya:2024tql}.

In the light-front quark model (LFQM)~\cite{Jaus:1991cy, Choi:2007se}, the radiative transition between ground state vector and pseudoscalar mesons is studied using a Gaussian wave function. Similarly, within a framework utilizing a harmonic oscillator (HO)~\cite{Peng:2012tr} and a modified version~\cite{Ke:2010vn}, the $M1$ radiative decays for higher excited states of quarkonium have also been analyzed. Additionally, the radiative transition has recently been studied using basis light-front quantization (BLFQ)~\cite{Li:2018uif}, where the light-front wave function (LFWF) is obtained by diagonalizing the light-front Hamiltonian directly. 
These developments suggest that a more realistic wave function of excited states is necessary to accurately describe experimental data.

In light-front models, achieving self-consistency using various current components and polarization vectors has been a long-standing issue~\cite{Cheng:1996if, Cheng:2003sm, Jaus:1999zv, Jaus:2002sv, Chang:2018zjq,Li:2017uug,Li:2021cwv,Keister:1993mg}, as the light-front zero mode may arise in some cases. 
In recent years, a significant progress has been made in developing a self-consistent LFQM that allows for the computation of physical observables independent of specific current components~\cite{Arifi:2023uqc,Arifi:2022qnd,Choi:2021mni,Choi:2013mda,Choi:2024ptc}. 
This approach is consistent with the Bakamjian-Thomas (BT) construction~\cite{Bakamjian:1953kh,Keister:1991sb}, where the interaction $V_{q\bar{q}}$ between quark and antiquark pairs is incorporated into the mass operator $M := M_0 + V_{q\bar{q}}.$
The construction of meson states in this framework ensures light-front energy conservation of four-momentum at the meson-quark vertex $P^- = p_q^- + p_{\bar{q}}^-,$ emphasizing the importance of using the invariant mass $M_0$ to satisfy the energy conservation relation $P^- = (M_0^2 + \bm{P}_\perp^2)/P^+.$
The physical mass of the meson $M$ that appears in the operator for computing observables should be replaced by the invariant mass $M_0$ accordingly. 
This approach has been proven to give self-consistent results for various observables such as decay constants~\cite{Arifi:2022qnd}, higher-twist distribution amplitudes~\cite{Arifi:2023uqc,Choi:2013mda}, and form factors~\cite{Choi:2021mni,Choi:2024ptc}, providing a more comprehensive understanding of hadron structure within the LFQM. 
However, the self-consistent analysis of the $M1$ radiative transition is yet to be done.

In this study, we aim to explore the properties of heavy quarkonia and $B_c$ mesons within the LFQM, with a particular focus on their excited states in the $S$ wave (pseudoscalar [$n^1S_0$] and vector [$n^3S_1$] mesons) to better understand their internal structure.
To achieve this, we first construct the meson wave functions by expanding the HO basis functions up to the $3S$ state, 
which is a crucial component of the LFQM. The expansion of HO basis functions has proven to be effective and successful in describing experimental constraints on mass spectra and decay constants, particularly for the first radial excitation ($2S$ state)\footnote{ {\color{blue} For simplicity, we have simplified the notation $n^{2S+1}L_J$ to $nL$ states, which covers both pseudoscalar and vector mesons.}}~\cite{Arifi:2022pal,Pandya:2024qoj}.

The present work serves as an extension to the second radial excitation, concentrating on heavy quarkonia and $B_c$ states, where $\psi(3S)$ and $\Upsilon(3S)$ states have been observed and can be used to constrain the model. 
We then provide predictions for $B_c$ mesons, which are relevant for the LHC experiment. 
Here we also adopt a screened confining potential that significantly impacts higher excited states. 
Note that the screened potential has been adopted in the nonrelativistic quark model~\cite{Li:2009zu, Deng:2016ktl,Hong:2022sht}, but the analysis using this potential in the LFQM to obtain the LFWF is not done. 
We then fit the model parameters to the mass spectra and decay constants, achieving reasonable agreement with experimental data and satisfying the observed hierarchy.
A preliminary result for charmonia using this framework has been reported in a conference proceeding~\cite{Ridwan:2024hyh}. 

Additionally, we extend the previous LFQM calculation of $M1$ radiative decays~\cite{Choi:2007se}, which used plus component ($\mu=+$) and transverse ($h=\pm1$) polarization, to include calculations with transverse component ($\mu=\perp$) and both longitudinal ($h=0$) and transverse ($h=\pm1$) polarizations. 
In the BLFQ analysis~\cite{Li:2018uif}, the transition form factor of the $M1$ transitions was found to differ depending on the polarization of the vector mesons. 
For the first time, we demonstrate that these observables, specifically the transition form factor remains consistent regardless of the current components and polarizations, 
due to the replacement $M \to M_0$, following the BT construction. 
We also find that the difference between longitudinal and transverse polarizations provides a measure of the anisotropy of the wave function. 
Given the isotropic nature of the HO basis functions used, 
the results for different polarizations become identical. 

For our numerical analysis, we present predictions for $M1$ radiative decays of heavy quarkonia and $B_c$ mesons up to the second radial excitation. 
We compare the predicted transition form factors for both \textit{allowed} $(n=n^\prime)$ and \textit{hindered} $(n\neq n^\prime)$ transitions, where $n$ is the principle quantum number.
We obtain that our numerical results for the coupling constants, decay widths, and branching ratios reasonably agree with available experimental and lattice QCD data, as well as other model calculations. 

This paper is organized as follows. In Sec.~\ref{sec:LFQM}, we present the formulation of the LFQM used in this study. 
We then explain the application of the LFQM to decay constants in Sec.~\ref{sec:constant} and to $M1$ radiative transitions in Sec.~\ref{sec:radiative}.
In Sec.~\ref{sec:result}, we show how we determine our model parameters and discuss our numerical results in comparison especially with experimental and lattice QCD data. 
We also address $M1$ radiative decays with different polarizations.
Finally, we conclude our present work in Sec.~\ref{sec:conclusion}. 
The wave function and mass formula are given explicitly in Appendix~\ref{app:WF}, followed by an explanation of the fitting procedure in Appendix~\ref{app:fitting}.
Some definitions used in this work, along with the derivations of the matrix elements for $M1$ radiative decays, are provided in Appendices~\ref{app:definitions} and~\ref{app:radiative}, respectively.

\section{Light-front quark model} 
\label{sec:LFQM}

In this section, we present the basic idea of the LFQM for describing the properties and structure of mesons~\cite{Choi:1997iq,Choi:2015ywa,Arifi:2022pal,Pandya:2024qoj}. 
The radial wave function is treated as a trial wave function using the HO basis function for the variational analysis of the QCD-motivated effective Hamiltonian, which saturates the Fock state expansion with the constituent quark and antiquark. 
Here, we consider heavy quarkonia and $B_c$ mesons up to the second radial excitation.

The meson system at rest is described as a bound system of effectively dressed valence quark and antiquark satisfying the eigenvalue equation of the QCD-motivated effective Hamiltonian, 
\begin{eqnarray}\label{eq:1}
H_{q\bar{q}} \ket{\Psi_{q\bar{q}}} = M_{q\bar{q}} \ket{\Psi_{q\bar{q}}},
\end{eqnarray}
where $M_{q\bar{q}}$ and $\Psi_{q\bar{q}}$ are the mass eigenvalue and eigenfunction of the $q\bar{q}$ meson state, respectively. We take the Hamiltonian $H_{q\bar{q}}$ in the quark-antiquark center of mass frame as 
$H_{q\bar{q}} = H_0 + V_{q\bar{q}}$ where
\begin{equation}
H_0 = \sqrt{m_q^2 + \bm{k}^2}  +  \sqrt{m_{\bar{q}}^2 + \bm{k}^2}
\end{equation}
is the kinetic energy part of the quark and antiquark with three-momentum $\bm{k}$. The effective inter-quark potential $V_{q\bar{q}}$ is given by
\begin{eqnarray}
V_{q\bar{q}} &=&  V_{\rm Conf} + V_{\rm Coul} + V_{\rm Hyp},
\end{eqnarray}
where $V_{\rm Conf}$ is the confining potential which includes the screened effect as~\cite{Li:2009zu, Deng:2016ktl, Hong:2022sht}
\begin{equation}
V_{\rm Conf} = a + \frac{b(1-{\rm e}^{-c r})}{c},
\end{equation}
with $a,b,$ and $c$ being parameters to be determined later.
The Coulomb and hyperfine interaction potentials stemming from the effective one-gluon exchanges for the $S$-wave mesons are written as
\begin{eqnarray} 
V_{\rm Coul} = -\frac{4\alpha_s}{3r}, \hspace{0.5cm}
V_{\rm Hyp} = \frac{32\pi \alpha_s (\bm{S}_q \cdot \bm{S}_{\bar{q}}) }{ 9 m_q m_{\bar{q}}} \delta^3(r).
\end{eqnarray}
We take the strong coupling $\alpha_s$ as a constant parameter and the values of the $\Braket{ \bm{S}_q \cdot \bm{S}_{\bar{q}} }$ are $1/4$ and $-3/4$ for the vector and pseudoscalar mesons, respectively.

The LFWF is represented by the Lorentz invariant internal variables 
\begin{eqnarray}
x_i &=& p^+_i /P^+, \\
\bm{k}_{\perp i} &=& \bm{p}_{\perp i} - x_i \bm{P}_{\perp},   
\end{eqnarray}
and helicity $\lambda_i$, where 
$P^\mu = (P^+,P^-,\bm{P}_\perp)$ is the four-momentum of the meson, and $(p^\mu_q, p^\mu_{\bar{q}})$ are the four-momentum of the quark and antiquark, respectively. 
This leads to the constraints $x_q + x_{\bar{q}} = 1$ and 
$\bm{k}_{\perp q} + \bm{k}_{\perp \bar{q}} = 0$.
Here we assign $x \equiv x_q$ with $\bm{k}_\perp \equiv \bm{k}_{\perp q}$.
The LFWF ($\Psi_{q{\bar q}}\equiv\Psi^{Jh}_{nS}$) of the $nS$ state pseudoscalar and vector mesons in the momentum space is then given by~\cite{Choi:1997iq,Choi:2015ywa,Arifi:2022pal,Pandya:2024qoj} 
\begin{eqnarray}\label{eq:LFWF}
		\Psi^{Jh}_{nS}(x, \bm{k}_{\bot},\lambda_i) = \Phi_{nS}(x, \bm{k}_\bot)
		\  \mathcal{R}^{Jh}_{\lambda_q\lambda_{\bar{q}}}(x, \bm{k}_\bot),
\end{eqnarray}
where $\Phi_{nS}$ is the radial wave function and $\mathcal{R}^{Jh}_{\lambda_q\lambda_{\bar{q}}}$ 
is the spin-orbit wave function that is obtained by the interaction-independent Melosh transformation~\cite{Melosh:1974cu} from the ordinary
spin-orbit wave function assigned by the quantum number $J^{PC}$.
The covariant forms of $\mathcal{R}^{Jh}_{\lambda_q\lambda_{\bar{q}}}$ for pseudoscalar $\mathcal{P}$ and vector $\mathcal{V}$ mesons are 
given by
\begin{eqnarray}
	\mathcal{R}^{Jh}_{\lambda_q\lambda_{\bar{q}}} &=&  \frac{1}{\sqrt{2} \tilde{M}_0} 
	\bar{u}_{\lambda_q}^{}(p_q) \Gamma_{\mathcal{P(V)}} v_{\lambda_{\bar{q}}}^{}(p_{\bar{q}}), 
\end{eqnarray}
where
\begin{eqnarray}
\Gamma_\mathcal{P} &=& \gamma_5 ,\\
\Gamma_\mathcal{V} &=& -\slashed{\epsilon}(h) + \frac{\epsilon \cdot (p_q-p_{\bar{q}})}{\mathcal{D}_{0} },
\end{eqnarray}
with $\tilde{M}_0 \equiv \sqrt{M_0^2 - (m_q -m_{\bar{q}})^2}$, $\mathcal{D}_{0} \equiv M_0 + m_q + m_{\bar{q}}$, and the boost-invariant mass of meson is given by
\begin{eqnarray}
	M_0^2 = \frac{\bm{k}_{\bot}^2 + m_q^2}{x}  + \frac{\bm{k}_{\bot}^2 + m_{\bar{q}}^2}{1-x}.
\end{eqnarray}
The polarization vectors $\epsilon^\mu(h)=(\epsilon^+, \epsilon^-,\bm{\epsilon}_{\perp})$ of the vector meson 
are given by
\begin{eqnarray}
\epsilon^\mu(\pm 1) &=& \left( 0, \frac{2}{P^+} \bm{\epsilon}_\perp(\pm) \cdot \bm{P}_\perp, \bm{\epsilon}_\perp(\pm)\right),
\nonumber\\
\epsilon^\mu(0) &=& \frac{1}{M_0}\left(P^+, \frac{-M^2_0 + \bm{P}^2_\perp}{P^+}, \bm{P}_\perp\right),
\end{eqnarray}
where $\bm{\epsilon}_\perp(\pm 1) = \mp \frac{1}{\sqrt{2}} \left( 1, \pm i \right).$
Note that $\mathcal{R}^{Jh}_{\lambda_q\lambda_{\bar{q}}}$ follows the orthonormal condition as 
\begin{eqnarray}
\sum_{\lambda_q,\lambda_{\bar{q}}} \braket{\mathcal{R}^{Jh}_{\lambda_q\lambda_{\bar{q}}} }{\mathcal{R}^{J^\prime h^\prime}_{\lambda_q\lambda_{\bar{q}}}} = \delta_{JJ^\prime}\delta_{h h^\prime}.
\end{eqnarray}
The equation above demonstrates the orthogonality between the pseudoscalar and vector mesons. The explicit form can be seen in Appendix~\ref{app:definitions}.

For the radial wave function of Eq.~\eqref{eq:LFWF}, 
we use the HO basis functions expanded up to the $3S$ state, \textit{i.e.}, $\Phi_i = R_{ij} \phi^{HO}_j$ and can be explicitly written as~\cite{Ridwan:2024hyh}
\begin{eqnarray}
\begin{pmatrix}
\Phi_{1S}  \\
\Phi_{2S}  \\
\Phi_{3S}  
\end{pmatrix}
&=&
\begin{pmatrix}
c_1^{1S} & c_2^{1S} & c_3^{1S} \\
c_1^{2S} & c_2^{2S} & c_3^{2S} \\
c_1^{3S} & c_2^{3S} & c_3^{3S} 
\end{pmatrix}
\begin{pmatrix}
\phi_{1S}^{\rm HO} \\
\phi_{2S}^{\rm HO} \\
\phi_{3S}^{\rm HO}  
\end{pmatrix} ,
\end{eqnarray}
where the $\bm{R}$ matrix contains
\begin{eqnarray}
\bm{R} = 
\begin{pmatrix}
    1 & 0 & 0 \\
    0 & c_{23} & s_{23} \\
    0 & -s_{23} & c_{23}  \\
\end{pmatrix} 
\begin{pmatrix}
    c_{13} & 0 & s_{13} \\
    0 & 1  & 0 \\
    -s_{13} & 0 & c_{13} \\
\end{pmatrix} 
\begin{pmatrix}
    c_{12} & s_{12} & 0 \\
    -s_{12} & c_{12}  & 0 \\
    0 & 0 & 1 \\
\end{pmatrix}
\nonumber\\
= 
\begin{pmatrix}
    c_{12}c_{13} & s_{12}c_{13} & s_{13} \\
    -s_{12}c_{23} - c_{12}s_{23}s_{13} & c_{12}c_{23} - s_{12}s_{23}s_{13} & s_{23}c_{13} \\
    s_{12}s_{23} - c_{12}c_{23}s_{13} & -c_{12}s_{23} - s_{12}c_{23}s_{13} & c_{23}c_{13} 
\end{pmatrix}, \nonumber\\
\end{eqnarray}
with $c_{ij}(s_{ij}) = \cos\theta_{ij}(\sin\theta_{ij})$. 
Note that the form of $\bm{R}$ is similar to the CKM matrix~\cite{Chau:1984fp}.
This form of $\bm{R}$ maintains the orthonormality of the wave functions 
\begin{eqnarray}
    \braket{\Phi_{nS}}{\Phi_{n^\prime S} } = \delta_{nn^\prime}.
\end{eqnarray}
This is an extension of the previous works~\cite{Arifi:2022pal,Syahbana:2024hkc} where only two HO basis functions up to the $2S$ state are considered.
For a more accurate approximation of the eigenstates of the Hamiltonian, one can use the Gaussian expansion method (GEM)~\cite{Hiyama:2003cu}. This method reproduces the asymptotic behavior at both short ($r \to 0$) and long ($r \to \infty$) ranges. A useful comparison between the wave functions obtained using GEM and those from a single Gaussian Ansatz can be found in Ref.~\cite{Arifi:2024mff}, highlighting their differences. The use of a few HO basis functions in this work provides a simple demonstration for explaining the mass gap and decay constant hierarchy, which is one of the focuses of this study.

In the momentum space, the HO basis functions up to the $3S$ state are expressed as
\begin{eqnarray}
	\phi_{1S}^\mathrm{HO} (\bm{k}) &=& \frac{1}{ \pi^{3/4}\beta^{3/2}} e^{-k^2/ 2\beta^2},\\
	\phi_{2S}^\mathrm{HO} (\bm{k}) &=& \frac{(2k^2 -3\beta^2)}{\sqrt{6} \pi^{3/4}\beta^{7/2}} e^{-k^2/ 2\beta^2},\\
 	\phi_{3S}^\mathrm{HO} (\bm{k}) &=& \frac{(15\beta^4 -20\beta^2k^2 + 4k^4)}{2\sqrt{30} \pi^{3/4}\beta^{11/2}}  e^{-k^2/ 2\beta^2}, \quad \quad 
\end{eqnarray}
with $k=|\bm{k}|$ and $k^2 = \bm{k}_\perp^2 + k_z^2$. The $\beta$ parameter, which controls the range of the wave function, is common for the $\phi_{nS}^\mathrm{HO}$ and, therefore, makes the basis functions orthogonal to each other.
Then, we perform a variable transformation $k_z \to x$ given by 
\begin{eqnarray}\label{eq:k_z}
    k_z = \left( x - \frac{1}{2} \right) M_0 + \frac{(m^2_{\bar{q}} -m^2_q)}{2M_0}.
\end{eqnarray} 
Thus, the radial wave function can be written as
\begin{eqnarray}
    \Phi_{nS}(x,\bm{k}_\bot) =  \sqrt{2(2\pi)^3} \sqrt{\frac{\partial k_z}{\partial x}} \Phi_{nS}(\bm{k}),
\end{eqnarray}
where the Jacobian factor 
\begin{equation}
\frac{\partial k_z}{\partial x} = \frac{M_0}{4x(1-x)} \left[ 1 - \frac{ (m_q^2 - m_{\bar{q}}^2)^2}{M_0^4} \right]
\end{equation}
should be included due to the variable transformation and is important to maintain the rotational symmetry~\cite{Arifi:2022qnd}.
We note that the LFWF follows the orthonormal condition as
\begin{eqnarray}
 \int \frac{\dd x\ \dd^2 \bm{k}_\bot}{2(2\pi)^3} \Psi_{nS}^{Jh\dagger} \Psi_{n^\prime S}^{J^\prime h^\prime}  = \delta_{JJ^\prime}\delta_{h h^\prime}\delta_{nn^\prime}.
\end{eqnarray}

The present LFQM analysis involves several parameters, including the constituent masses of charm and bottom quarks $(m_c, m_b)$, the potential parameters $(a, b, c, \alpha_s)$, the wave function parameter $\beta_{q\bar{q}}$ for each $(q\bar{q})$ content, and mixing parameters $\theta_{ij}$, which are assumed to be common and are found to be a good approximation~\cite{Arifi:2022pal}.
To determine the values of these parameters, we use the variational principle which involves the condition
\begin{eqnarray}
	\frac{\partial \bra{\Psi_{q\bar q}}  (H_0+V_\mathrm{conf}+V_\mathrm{coul})  \ket{\Psi_{q\bar q}} }{\partial \beta_{q\bar{q}}} = 0,
\end{eqnarray}
where the spin-spin interaction is treated perturbatively for simplicity.
The variational principle is essential for constraining the range of the wave functions. 
The model parameters are then fitted to the mass spectra and decay constants, which will be explained later in Sec.~\ref{sec:parameter}. 
With the optimized parameters, we predict the $M1$ radiative transitions between various states.

\section{Decay constants} 
\label{sec:constant}

In this section, we present the application of the LFQM for the decay constant, which is one of the fundamental properties of mesons and is sensitive to the short-distance part of the wave function.
The decay constant of a pseudoscalar meson $f_\mathcal{P}$ and a vector meson $f_\mathcal{V}$ with a four-momentum $P^\mu$ are defined by
\begin{eqnarray}
	\bra{0} \bar{q}(0) \gamma^\mu \gamma^5 q(0) \ket{\mathcal{P}(P)} &=& i f_\mathcal{P} P^\mu,\\
	\bra{0} \bar{q}(0) \gamma^\mu q(0) \ket{\mathcal{V}(P,h)} &=& f_\mathcal{V} M \epsilon^\mu,
\end{eqnarray}
where $M$ and $\epsilon^\mu (P,h)$ are the mass and the polarization vector of the vector meson, respectively. 

In the standard LFQM, the decay constant can be calculated as
\begin{eqnarray}
f_\mathcal{P(V)} &=& \sqrt{3} \int_0^1 \mathrm{d}x \int \frac{\mathrm{d}^2 \bm{k}_\bot}{16\pi^3}\ \frac{\Phi(x,\bm{k}_\perp) }{\mathcal{G}_\mathcal{P(V)}^\mu} \nonumber\\
&&\times \sum_{\lambda_1, \lambda_2} \mathcal{R}_{\lambda_1 \lambda_2}^{Jh} 
\left[\frac{\bar{v}_{\lambda_2}(p_2)}{\sqrt{x_2}}
   \Gamma_\mathcal{P(V)} \frac{u_{\lambda_1}(p_1)}{\sqrt{x_1}}\right], \quad
\end{eqnarray}
where $\Gamma_\mathcal{P}=\gamma^\mu\gamma^5$ and $\Gamma_\mathcal{V}=\gamma^\mu$, and $\mathcal{G}_\mathcal{P}^\mu=iP^\mu$ and $\mathcal{G}_\mathcal{V}^\mu=M\epsilon^\mu$. 
Note that the replacement $M\to M_0$ should be carried out when obtaining the operator to get a self-consistent result.
The explicit form of $f_\mathcal{P(V)}$ is given by~\cite{Arifi:2022pal}
\begin{eqnarray}\label{eq:constant}
f_\mathcal{P(V)} &=& \sqrt{6} \int \frac{\dd x\ \dd^2 \bm{k}_\bot}{2(2\pi)^3}  
	\frac{ {\Phi}(x, \bm{k}_\bot) }{\sqrt{\mathcal{A}^2 + \bm{k}_\bot^2}} ~\mathcal{O}_\mathcal{P(V)}^\mu(h), \quad \quad 
\end{eqnarray}
where the operator for the pseudoscalar mesons reads
\begin{eqnarray}
        \mathcal{O}_\mathcal{P}^+ &=& 2\mathcal{A},
\end{eqnarray}
and those for the vector mesons with different polarizations read
\begin{eqnarray}
    \mathcal{O}_\mathcal{V}^+(0) &=& 2\left(\mathcal{A} + \frac{2 \bm{k}_\bot^2}{\mathcal{D}_{0}}\right),\\
    \mathcal{O}_\mathcal{V}^\perp(\pm1) &=& \left( \frac{\mathcal{A}^2+\bm{k}_\perp^2}{x(1-x)M_0} - \frac{2\bm{k}_\perp^2}{\mathcal{D}_0}\right),
\end{eqnarray}
with $\mathcal{A}=xm_{\bar{q}}+(1-x)m_q$.
In the LFQM, it has been shown that the decay constants can be obtained consistently and yield the same results using various currents, polarizations, and reference frames~\cite{Arifi:2022qnd,Arifi:2023uqc}. 
The difference between the operators of decay constant computed using the longitudinal and transverse polarizations can be written analytically as
\begin{eqnarray}
 \Delta \mathcal{O}_{\mathcal{V}} &=& \mathcal{O}_{\mathcal{V}}^\perp(\pm1) - \mathcal{O}_{\mathcal{V}}^+(0) \nonumber\\
 &=& -\frac{2}{\mathcal{D}_0}(\bm{k}_\perp^2-2k_z^2) + \frac{2(m_q-m_{\bar{q}})}{M_0}k_z,
\end{eqnarray}
which is vanishing after evaluating the momentum integration. 
The second term clearly vanishes in the equal-mass case ($m_q=m_{\bar{q}}$) and involves an odd term in $k_z$, which also vanishes after integration over $\dd x$.
However, it will not vanish unless the wave function is rotationally symmetric, which leads to 
\begin{eqnarray}
    \expval{k_\perp^2}=\expval{2k_z^2}.
\end{eqnarray}

\begin{figure}[t]
	\centering
	\includegraphics[width=0.9\columnwidth]{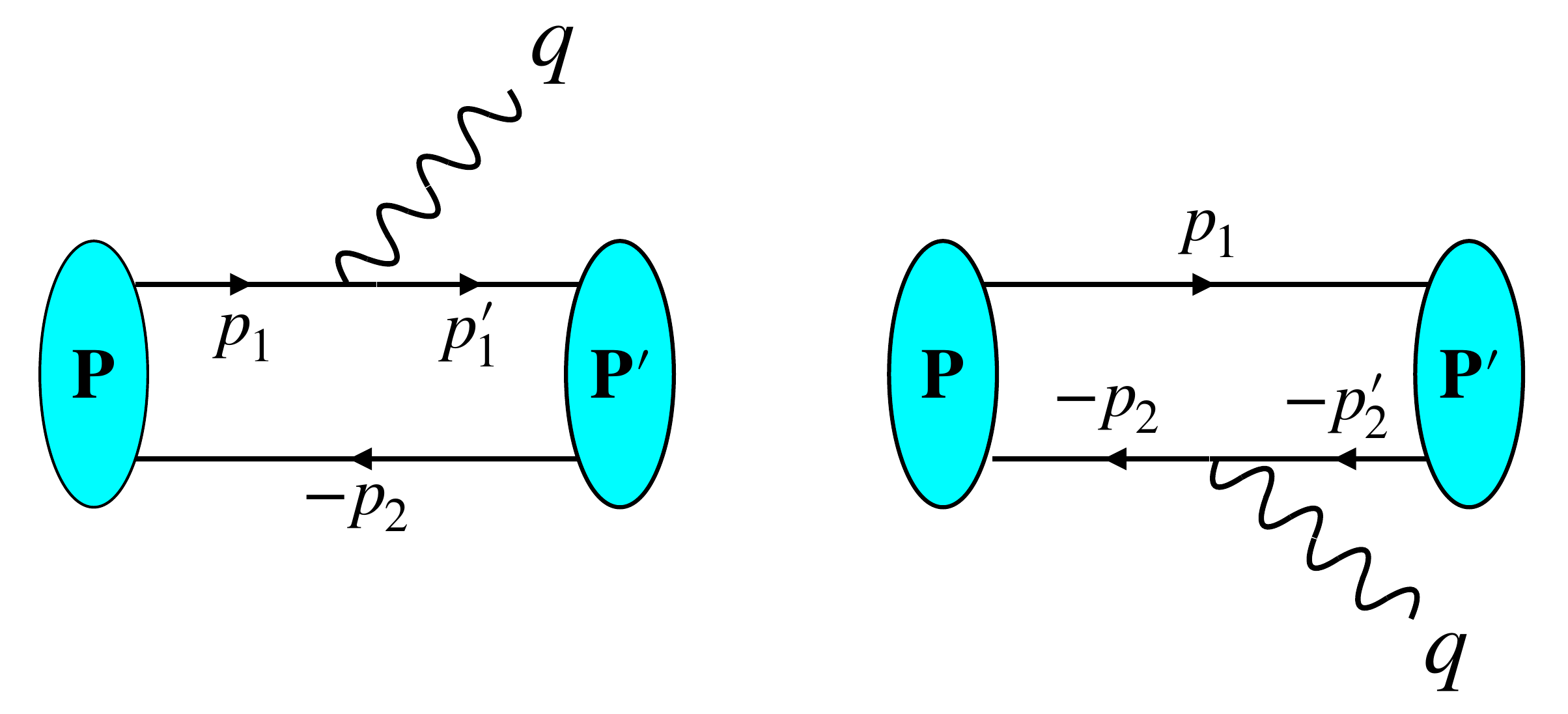}
	\caption{\label{fig:RD}  The Feynman diagrams for the $M1$ radiative transitions $\mathcal{V}(\mathcal{P}) \to \mathcal{P}(\mathcal{V}) \gamma$, where the photon couples to either the quark or antiquark, which are added coherently when computing the decay width.}
\end{figure}

\begin{table}[b]
	\centering
 	\begin{ruledtabular}
  		\renewcommand{\arraystretch}{2.2}
	\caption{The right-hand side of Eq.~\eqref{eq:rad_M1} for various components of the current and polarizations. We define $ q^{R(L)} = q_x \pm i q_y $ and have used the $ q^+ = 0 $ frame. Note that we cannot extract the form factor using $\mu=R(L)$ with $h=+1(-1)$ and $\bm{P}_{\perp} = 0$ since $\mathcal{G}^\mu_h$ vanishes, similar to the case when using $\mu=+$ and $h=0$. }
	\label{tab:tensor}
	\begin{tabular}{c c c c}
		$\mu$ & $\epsilon(h)$ & $\mathcal{G}^\mu_h(\bm{P}_\perp\neq 0)$ &  $\mathcal{G}^\mu_h(\bm{P}_\perp = 0)$ \\ \hline
            + & $\epsilon(0)$ & 0  & 0\\
             & $\epsilon(+1)$ & $\dfrac{eP^+q^R}{\sqrt{2}} $ & $\dfrac{eP^+q^R}{\sqrt{2}} $\\
             & $\epsilon(-1)$ & $\dfrac{eP^+q^L}{\sqrt{2}}$ & $\dfrac{eP^+q^L}{\sqrt{2}}$\\
            $R$ & $\epsilon(0)$ & $-e M q^R $ & $-e M q^R $\\
             & $\epsilon(+1)$ & $\dfrac{-e P^R q^R}{\sqrt{2}}$ & 0\\
             & $\epsilon(-1)$ & $\dfrac{e\left(q^- P^+ + P^L q^R \right)}{\sqrt{2}} $ & $\dfrac{e q^- P^+}{\sqrt{2}} $\\
            $L$ & $\epsilon(0)$ & $e M q^L $  & $e M q^L $ \\
             & $\epsilon(+1)$ & $\dfrac{e\left(q^- P^+ + P^R q^L \right)}{\sqrt{2}} $ & $\dfrac{e q^- P^+}{\sqrt{2}} $ \\
            & $\epsilon(-1)$ & $\dfrac{-eP^L q^L}{\sqrt{2}}$ & 0 \\
        \end{tabular}
     \renewcommand{\arraystretch}{1}
   \end{ruledtabular}
\end{table}

\section{$M1$ radiative transitions} 
\label{sec:radiative}

In this study, we focus on analyzing $M1$ radiative transitions in heavy quarkonia and $B_c$ mesons. These $M1$ transitions are characterized by changes in the spin quantum number ($\Delta S = 1$) while leaving the orbital angular momentum unchanged ($\Delta L=0$). Specifically, we examine transitions between vector (pseudoscalar) mesons and pseudoscalar (vector) mesons.

In the LFQM, there are roughly two approaches for computing radiative decays. The first is the "standard LFQM," where the Melosh transformation is used to derive the spin-orbit wave function~\cite{Jaus:1991cy}. 
The second is the "covariant LFQM," where the trace technique is used to evaluate the matrix elements derived from one-loop Feynman diagrams, ensuring Lorentz covariance in the light-front calculations~\cite{Jaus:1999zv}. 
Despite the different techniques, the results from the "covariant LFQM" will lead to the same outcomes as those in the "standard LFQM" by performing the "type-II" replacement, as proposed in Ref.~\cite{Choi:2013mda}. 
This mapping has been discussed recently to check the self-consistency of the model. In this work, we will use the standard LFQM to describe the radiative decay.

Within the LFQM, the Feynman diagram for the $\mathcal{V}(\mathcal{P}) \to \mathcal{P}(\mathcal{V}) \gamma$ process is shown in Fig.~\ref{fig:RD}. 
This diagram illustrates the transition of a vector meson into a pseudoscalar meson, and vice versa. 
The initial state emits a photon, mediated by a quark loop.
For the $M1$ transition of vector meson $\mathcal{V} \to \mathcal{P} \gamma$, the $F_{\mathcal{VP}}(Q^2)$ form factor is defined as~\cite{Choi:2007se}
\begin{eqnarray}
\label{eq:rad_M1}
\bra{\mathcal{P}(P^\prime)} J_{\rm em}^\mu(0)\ket{\mathcal{V}(P,h)} = ie \varepsilon^{\mu \nu \rho\sigma} \epsilon_\nu q_\rho P_\sigma F_\mathcal{VP}(Q^2),\quad 
\end{eqnarray}
where the antisymmetric tensor $ \varepsilon^{\mu \nu \rho\sigma} $ assures electromagnetic gauge invariance and $q = P-P^\prime$ is the four-momentum of the virtual photon. 
The  $F_{\mathcal{VP}}(Q^2)$ can be obtained in the Drell-Yan-West frame, where $q^+ = 0$. In this frame, the momentum transfer can be written as $q^2 = q^+q^- - \bm{q}_\perp^2 = -\bm{q}_\perp^2 \equiv -Q^2$.

\begin{table*}[t]
	\centering
 	\begin{ruledtabular}
  		\renewcommand{\arraystretch}{1.5}
	\caption{Operators for the radiative transitions obtained using the plus ($\mu=+$) and transverse [$\mu=R(L)$] components with both longitudinal ($h=0$) and transverse ($h=\pm1$) polarizations in the LFQM. We have substituted $M^{(\prime)}$ with $M_0^{(\prime)}$ to derive the operator associated with the transverse component [$\mu = R(L)$], ensuring self-consistent results. }
	\label{tab:operator}
	\begin{tabular}{c c l }
		$\mu$ & $\epsilon(h)$ & $\mathcal{O}$ \\ \hline
            \multirow{2}{*}{$+$} & \multirow{2}{*}{$\epsilon(0)$}  & \multirow{2}{*}{\dots} \\
                & & \\
           \multirow{2}{*}{$+$}   & \multirow{2}{*}{$\epsilon(\pm1)$} & \multirow{2}{*}{$2(1-x)\left[\mathcal{A} + \dfrac{2}{\mathcal{D}_0} 
	\biggl(\bm{k}^2_{\perp} - \dfrac{(\bm{k}_{\perp} \cdot \bm{q}_{\perp})^2}{\bm{q}^2_{\perp}} \right)\biggr] $} \\ 
                & & \\ \hline
            \multirow{2}{*}{$R(L)$}   & \multirow{2}{*}{$\epsilon(0)$} & \multirow{2}{*}{$\dfrac{1}{x M_0}\biggl\{ \mathcal{A}\left(\mathcal{A}+\dfrac{2\bm{k}_\perp^2}{\mathcal{D}_0}\right) + \dfrac{\mathcal{M}}{ \mathcal{D}_0} \left[(1-2x)\bm{k}_\perp^2  +(1-x) \left( (\bm{k}_\perp\cdot \bm{q}_\perp) -\dfrac{2(\bm{k}_\perp\cdot \bm{q}_\perp)^2}{\bm{q}_\perp^2}  \right)\right] \biggr\}$} \\  
                & & \\
            \multirow{2}{*}{$R(L)$}  & \multirow{2}{*}{$\epsilon(-1)[\epsilon(+1)]$}  & \multirow{2}{*}{$\dfrac{2}{x(M_0^2-M_0^{\prime 2} -\bm{q}_\perp^2)} \left[ (\bm{k}_\perp\cdot\bm{q}_\perp) \left(\mathcal{A} + \dfrac{x \bm{k}_{\perp}^2}{\mathcal{D}_0} - \dfrac{\mathcal{A}\mathcal{M}_1}{\mathcal{D}_0} \right) + (1-x)(\bm{k}_\perp\cdot\bm{q}_\perp -\bm{q}_\perp^2) \left(\mathcal{A} + \dfrac{\bm{k}_{\perp}^2}{\mathcal{D}_0}\right) \right]$ }\\
             & & \\
       \multirow{2}{*}{$R(L)$}  & \multirow{2}{*}{$\epsilon(+1)[\epsilon(-1)]$} & \multirow{2}{*}{$2(1-x)\left[\mathcal{A} + \dfrac{2}{\mathcal{D}_0} 
	\biggl(\bm{k}^2_{\perp} - \dfrac{(\bm{k}_{\perp} \cdot \bm{q}_{\perp})^2}{\bm{q}^2_{\perp}} \right)\biggr] $} \\ 
 & & \\
     \end{tabular}
     		\renewcommand{\arraystretch}{1}
    \end{ruledtabular}
\end{table*}

The Lorentz structure in the right-hand side of Eq.~\eqref{eq:rad_M1} denoted as 
$\mathcal{G}_h^\mu=ie \varepsilon^{\mu \nu \rho\sigma} \epsilon_\nu q_\rho P_\sigma$ can be computed using different current components and polarizations. 
By evaluating the antisymmetric tensor, we obtain $\mathcal{G}_h^\mu$, as summarized in Table~\ref{tab:tensor}.
Traditionally, we use the plus component ($\mu = +$) with transverse polarization ($h = \pm 1$), as in Ref.~\cite{Choi:2007se}. 
Note that for the longitudinal polarization ($h = 0$), $\mathcal{G}_0^+$ vanishes, making it impossible to extract the form factor. 
In this work, we calculate the transverse component ($\mu = \perp$), which translates to [$\mu = R(L)$] with $R(L)=x\pm i y$, for both longitudinal ($h = 0$) and transverse ($h = \pm 1$) polarizations.
Here, we set $\bm{P}_\perp = 0$ for simplicity, except for the cases with $\mu = R(L)$ and $h = +1(-1)$. 
In these cases, we initially use a nonzero $\bm{P}_\perp$ frame because $\mathcal{G}_{+1}^R$ and $\mathcal{G}_{-1}^L$ vanish if $\bm{P}_\perp = 0$, as shown in Table~\ref{tab:tensor}. 
However, we set $\bm{P}_\perp \to 0$ when presenting the operator, 
which leads to the same formula as those for the plus component ($\mu=+$), as will be discussed later.

Now, let us compute the left-hand side of Eq.~\eqref{eq:rad_M1} in the LFQM. 
First, we examine the momentum conservation for $\mathcal{V}(P) \to \mathcal{P}(P^\prime) + \gamma(q)$ transitions in the $q^+ = 0$ frame. 
We then obtain
\begin{eqnarray}
    P &=& \left(P^+,\frac{M^2+\bm{P}_\perp^2}{P^+},\bm{P}_\perp\right),\\
    P^\prime &=& \left(P^+,\frac{M^{\prime 2}+\bm{P}_\perp^{\prime 2} }{P^+},\bm{P}_\perp^\prime\right),\\
    q &=& \left(0, \frac{M^2 - M^{\prime 2} + \bm{P}_\perp^2-\bm{P}_\perp^{\prime 2} }{P^+},\bm{q}_\perp\right),
\end{eqnarray}
where $\bm{P}_\perp^\prime=\bm{P}_\perp-\bm{q}_\perp$.
In the quark level, the initial and final quark momenta are denoted as
\begin{eqnarray}
\mathrm{Initial\ state} &:&
\begin{cases}
    p_1^+ = x_1 P^+,\quad \bm{p}_{1\perp} = x_1 \bm{P}_\perp + \bm{k}_\perp, \\
    p_2^+ = x_2 P^+,\quad \bm{p}_{2\perp} = x_2 \bm{P}_\perp - \bm{k}_\perp.
\end{cases} \\
\mathrm{Final\ state} &:&
\begin{cases}
    p_1^{\prime +} = x_1 P^+, \quad \bm{p}_{1\perp}^\prime = x_1 \bm{P}^\prime_\perp + \bm{k}_\perp^\prime, \\
    p_2^{\prime +} = x_2 P^+, \quad \bm{p}_{2\perp}^\prime = x_2 \bm{P}^\prime_\perp - \bm{k}_\perp^\prime.
\end{cases}\quad
\end{eqnarray}
In the $q^+=0$ frame, the spectator quark requires $p_2^+ = p_2^{\prime +}$ and $\bm{p}_{2\perp} = \bm{p}_{2\perp}^\prime$, while for the struck quark, we have $p_1^+ = p_1^{\prime +}$ and  $\bm{p}_{1\perp} = \bm{p}_{1\perp}^\prime + \bm{q}_\perp$.
From these quark momentum conservation, one can obtain
\begin{eqnarray} \label{eq:kprime}
    \bm{k}_\perp^\prime = \bm{k}_\perp - (1-x) \bm{q}_\perp.
\end{eqnarray}

The matrix element $\mathcal{J}_h^\mu\equiv\bra{\mathcal{P}(P^\prime)} J_{\text{em}}^{\mu} \ket{\mathcal{V}(P,h)}$ can be written by the convolution formula of the initial and final state LFWFs as
\begin{eqnarray}
\mathcal{J}_h^\mu  &=& \sum_{\lambda,\lambda^\prime,\bar{\lambda},j} \mel{\Psi_{\lambda^\prime \bar{\lambda}}^{00\dagger} }{\frac{\bar{u}_{\lambda^\prime}(p_1^\prime) }{\sqrt{x}} e e_q^j\gamma^{\mu} 
\frac{u_{\lambda}(p_1) }{\sqrt{x}} }{\Psi_{\lambda \bar{\lambda}}^{1h}}\nonumber\\
&=& \sum_j e e_q^j \int \frac{\dd x\ \dd^2\bm{k}_\perp}{2(2\pi)^3}  \Phi(x, \bm{k}_\perp^\prime) \Phi(x,\bm{k}_\perp) \nonumber\\
& &\times \sum_{\lambda,\lambda^\prime,\bar{\lambda}} \mathcal{R}_{\lambda^\prime \bar{\lambda}}^{00\dagger}(x, \bm{k}_\perp^\prime) \frac{\bar{u}_{\lambda^\prime}(p_1^\prime) }{\sqrt{x}} \gamma^{\mu} 
\frac{u_{\lambda}(p_1) }{\sqrt{x}} \mathcal{R}_{\lambda \bar{\lambda}}^{1h}(x, \bm{k}_\perp),\nonumber\\
\end{eqnarray}
where $e_q^j$ is the electric charge for $j$-th quark flavor [$e_c (e_b) = 2/3 (-1/3)$].
Note that one should use $\bm{k}_\perp^{\prime}$ defined in Eq.~\eqref{eq:kprime} in the final state LFWF.
For instance, the invariant mass in the final state becomes
\begin{eqnarray}
    M_0^\prime(x,\bm{k}_\perp^{\prime}) = \frac{m_q^2 + \bm{k}_\perp^{\prime 2}}{x} +  \frac{m_{\bar{q}}^2 + \bm{k}_\perp^{\prime 2}}{1-x}.
\end{eqnarray}

To begin with, we derive the matrix elements not only for $\mathcal{J}_{\pm1}^+$, but also for $\mathcal{J}_{0}^{R(L)}$ and $\mathcal{J}_{\pm1}^{R(L)}$. 
By matching the left- and right-hand sides of Eq.~\eqref{eq:rad_M1},
the form factor can be computed as
\begin{eqnarray}
    F_\mathcal{VP}(Q^2) = \frac{\mathcal{J}_h^\mu}{\mathcal{G}_h^\mu}.
\end{eqnarray}
The form factor is then given by
\begin{eqnarray}
F_\mathcal{VP}(Q^2) = e_q I^{\mu}_h(m_q,m_{\bar{q}},Q^2) + e_{\bar{q}} I^{\mu}_h(m_{\bar{q}},m_q,Q^2),
\end{eqnarray}
which represents a process where the photon couples to the quark and antiquark, respectively, as illustrated in Fig.~\ref{fig:RD}.
The one-loop integral $I^{\mu}_{h}$ is given by
\begin{eqnarray}
\label{eq:operator}
 I^{\mu}_{h} = \int \frac{\dd x\dd^2 \bm{k}_\perp }{2(2\pi)^3} \frac{\Phi(x,\bm{k}_\perp^\prime) \Phi(x,\bm{k}_\perp)}{\sqrt{\mathcal{A}^2+\bm{k}_\perp^{\prime 2}} \sqrt{\mathcal{A}^2+\bm{k}_\perp^2}  } \mathcal{O}_{\mathcal{VP}\gamma}^\mu(h),
\end{eqnarray}
where the operator derived using various current components and polarizations are presented in Table~\ref{tab:operator}.
The detailed derivations are provided in Appendix~\ref{app:radiative}. We have replaced $M^{(\prime)}$ with $M_0^{(\prime)}$ to obtain the operator associated with the transverse components [$\mu = R(L)$] to achieve self-consistent results. It is also noteworthy that the operators for the transverse component [$\mu = R(L)$] with transverse polarizations ($h = \pm 1$) have two different forms, one of which coincides with those for the plus component ($\mu = +$).

\begin{table*}[t]
	\begin{ruledtabular}
		\renewcommand{\arraystretch}{1.3}
		\caption{Model parameters and their uncertainties, determined from fitting to mass spectra and decay constants using iMinuit~\cite{iminuit,James:1975dr}. The parameters include the constituent quark masses ($m_c$ and $m_b$), the potential parameters $a$ and $c$ (all in units of GeV), the dimensionless strong coupling constant $\alpha_s$, and the mixing angles $\theta_{12}$, $\theta_{13}$, and $\theta_{23}$. The mixing angles control the expansion coefficients, with the assumption that $\theta_{13} = \theta_{23}$. The string tension is fixed to the known value of $b = 0.18$ GeV$^2$, whereas the parameters $\beta_{q\bar{q}}$ are obtained from variational principle.}
		\label{tab:parameter}
		\begin{tabular}{cccccccc|ccc}
			$\theta_{12}$ & $\theta_{13}=\theta_{23}$ & $m_c$ & $m_b$ & $a$ & $b$ & $c$ & $\alpha_s$ &
			$\beta_{c\bar{c}}$ & $\beta_{c\bar{b}}$ & 
			$\beta_{b\bar{b}}$\\ \hline 
			 $12.12(30)^\circ$ & $8.44(14)^\circ$ & $1.61(4)$ & $4.97(4)$ & $-0.411(67)$ & $0.18$ & $0.027(12)$ & $0.402(4)$ & $0.542(8)$ & $0.702(9)$ & $1.060(11)$ \\
		\end{tabular}
		\renewcommand{\arraystretch}{1}
	\end{ruledtabular}
\end{table*}

For the physical process of the $M1$ radiative transition emitting real photon, we should take the $Q^2 \to 0$ limit to extract the coupling constant $g_{\mathcal{VP}\gamma} \equiv F_{\mathcal{VP}}(Q^2\to 0)$. 
In this limit, some operators also have the analytic formulas expressed as
\begin{eqnarray}
     \mathcal{O}_{\mathcal{VP}\gamma}^+(\pm1) &=& 2(1-x)\left(\mathcal{A} + \frac{\bm{k}^2_{\perp}}{\mathcal{D}_0}\right), \\
 \mathcal{O}_{\mathcal{VP}\gamma}^{R(L)}(0) &=& \frac{\mathcal{A}}{xM_0}\left(\mathcal{A} +\frac{2\bm{k}^2_{\perp}}{\mathcal{D}_0}\right). 
\end{eqnarray}
The difference of the two operators is expressed as 
\begin{eqnarray}
 \Delta \mathcal{O}_{\mathcal{VP}\gamma} &=& \mathcal{O}_{\mathcal{VP}\gamma}^+(\pm1) - \mathcal{O}_{\mathcal{VP}\gamma}^{R(L)}(0) \nonumber\\
  &=&  \frac{2}{\mathcal{D}_0}\left[ \frac{2k_z^2\bm{k}_\perp^2}{\bm{k}_\perp^2 + m^2} + \bm{k}_\perp^2 -2k_z^2\right],\quad 
\end{eqnarray}
where we have simplified the formula for the equal-mass case $m_q=m_{\bar{q}}=m$ to better illustrate the difference.
Again, the $k_\perp^2$ and $2k_z^2$ terms appear in the operator difference which measures the anisotropy of the wave function. 
The additional factor of $k_z^2\bm{k}_\perp^2$ appears because it is a transition process.

The partial decay width for the $\mathcal{V} \to \mathcal{P}\gamma$ transition is then computed as
\begin{eqnarray}
\label{eq:DWtheo}
	\Gamma(\mathcal{V}\to \mathcal{P}\gamma) = \frac{\alpha_\mathrm{em}}{(2J_{\mathcal{V}}+1)} g^2_{\mathcal{VP}\gamma} k^3_\gamma,
\end{eqnarray}
where $\alpha_\mathrm{em}=1/137$ is the EM fine-structure constant and 
\begin{eqnarray} \label{eq:k_photon}
	k_\gamma = \frac{(M_\mathcal{V}^2- M_\mathcal{P}^2)}{2M_\mathcal{V}}
\end{eqnarray}
is the kinematically allowed three-momentum $k_\gamma=\abs{\bm{k}_\gamma}$ of the outgoing real photon.
For the $\mathcal{P}\to \mathcal{V} \gamma$ transition, we need to modify $k_\gamma$ by exchanging $M_\mathcal{V}$ and $M_\mathcal{P}$ in Eq.~\eqref{eq:k_photon}.
In addition, the spin factor $J_\mathcal{V}=1$ should be replaced by $J_\mathcal{P}=0$ in Eq.~\eqref{eq:DWtheo}. 
After obtaining the decay width, we can calculate the branching ratio, defined as
\begin{equation}
\label{Eq:Br}
	\mathrm{Br}(\mathcal{V}\to \mathcal{P}\gamma) = \frac{\Gamma(\mathcal{V}\to \mathcal{P}\gamma)}{\Gamma_{\rm Total}},
\end{equation}
where the total decay width $\Gamma_{\rm Total}$ for each meson is taken from particle data group (PDG)~\cite{ParticleDataGroup:2024cfk}.

\section{Numerical results and discussion}
\label{sec:result}

In this section, we first discuss the model parameters, followed by a discussion of the LFWFs and various observables, including mass spectra, decay constants, and $M1$ radiative transitions. 
Comparisons with experimental data, Lattice QCD results, and other models are also presented.

\begin{figure}[t]
	\centering
	\includegraphics[width=1\columnwidth]{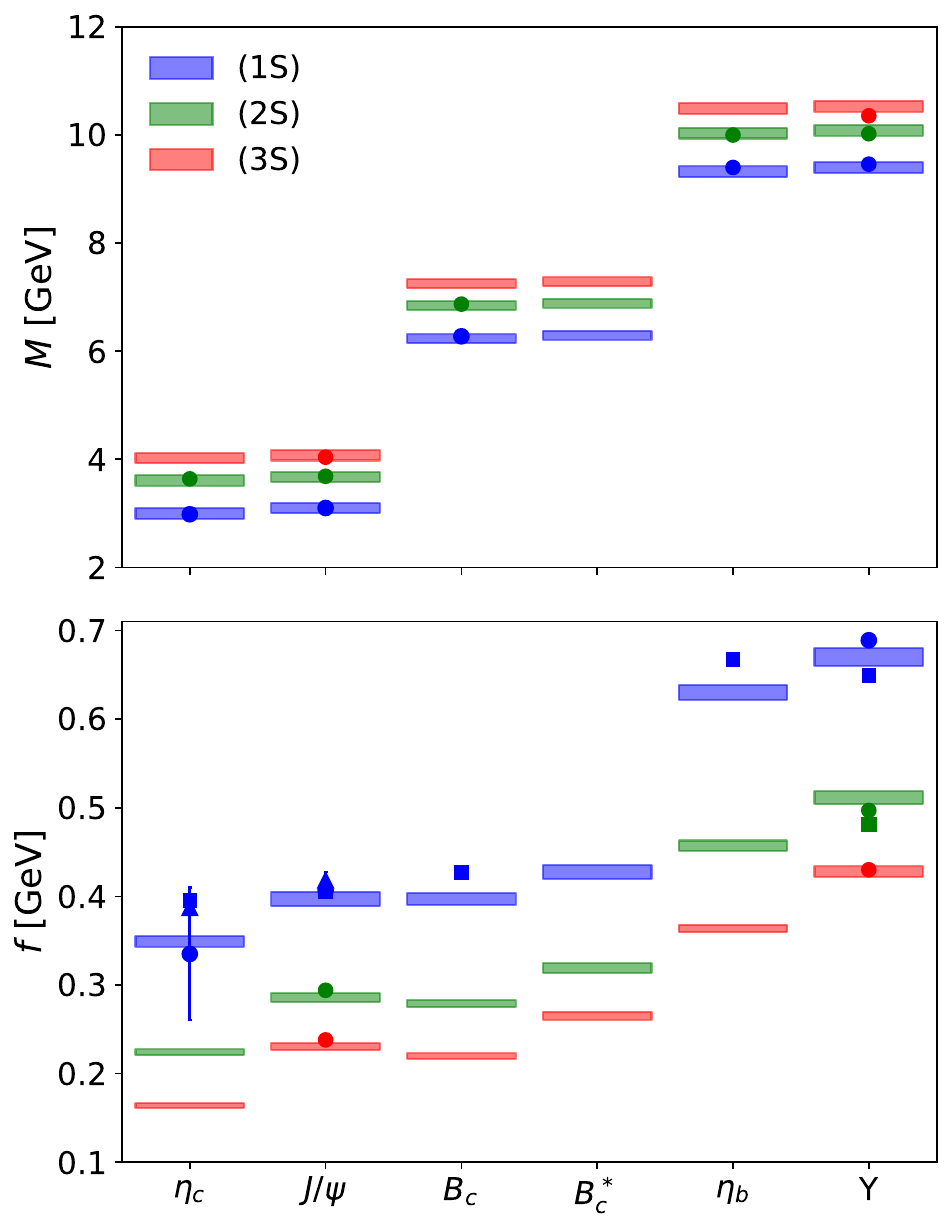}
\caption{[\textbf{Upper panel}] Mass spectra and [\textbf{Lower panel}] decay constants of $B_c$ and heavy quarkonia. Comparison of our model with experimental data (black circle), lattice QCD results (black square), and sum rule (black triangle) is presented. It shows that decreasing decay constants for the excited states can be reproduced in the present LFQM by expanding the HO basis functions. }
    \label{fig:massandDC}
\end{figure}

\subsection{Model parameters}
\label{sec:parameter}

Before discussing the LFWFs and observables, it is essential to determine the model parameters, which are constrained by variational analysis~\cite{Choi:2015ywa}. 
To simplify the analysis, we fix the well-known string tension at $b = 0.18$ GeV$^2$ and assume $\theta_{13} = \theta_{23}$. 
The model parameters are obtained by simultaneously fitting the available experimental data of both mass spectra and decay constants, following a similar procedure as in Ref.~\cite{Arifi:2024mff}.
The fitting process is performed by minimizing the $\chi^2$ defined by
\begin{eqnarray}\label{eq:chi2}
    \chi^2 = \sum_i\frac{(O_{i}^\mathrm{expt}-O^\mathrm{mod}_{i})^2}{(\sigma^\mathrm{mod}_{i})^2},
\end{eqnarray}
using iMinuit package~\cite{iminuit,James:1975dr}, where we define $\sigma^\mathrm{mod}=r O_{i}^\mathrm{expt}$ with $r=0.5\%$ and $1\%$ for the mass spectra and decay constants, respectively. 
The $\sigma^\mathrm{mod}$ will reduce the fitting bias caused by the different precisions of experimental data. 
The uncertainty of the parameters is obtained using the Hesse method in the iMinuit package, and the uncertainty propagation is computed using the Monte Carlo method.
In addition to that statistical uncertainty from the fit, we note that excited states are more sensitive to the choice of trial wave function, which leads to greater systematic uncertainty compared to the ground state, see Fig.~2 of Ref.~\cite{Arifi:2022pal}.
It is worth noting that we cannot fix the model parameters by considering only the mass spectra or the decay constants; both are crucial in constraining the parameters. Therefore, in the fits, the parameters are adjusted automatically to achieve a reasonable agreement with both observables. 
Further details on the fitting procedure and uncertainty propagation are discussed in Appendix~\ref{app:fitting}.

\begin{figure}[t]
	\centering
	\includegraphics[width=1\columnwidth]{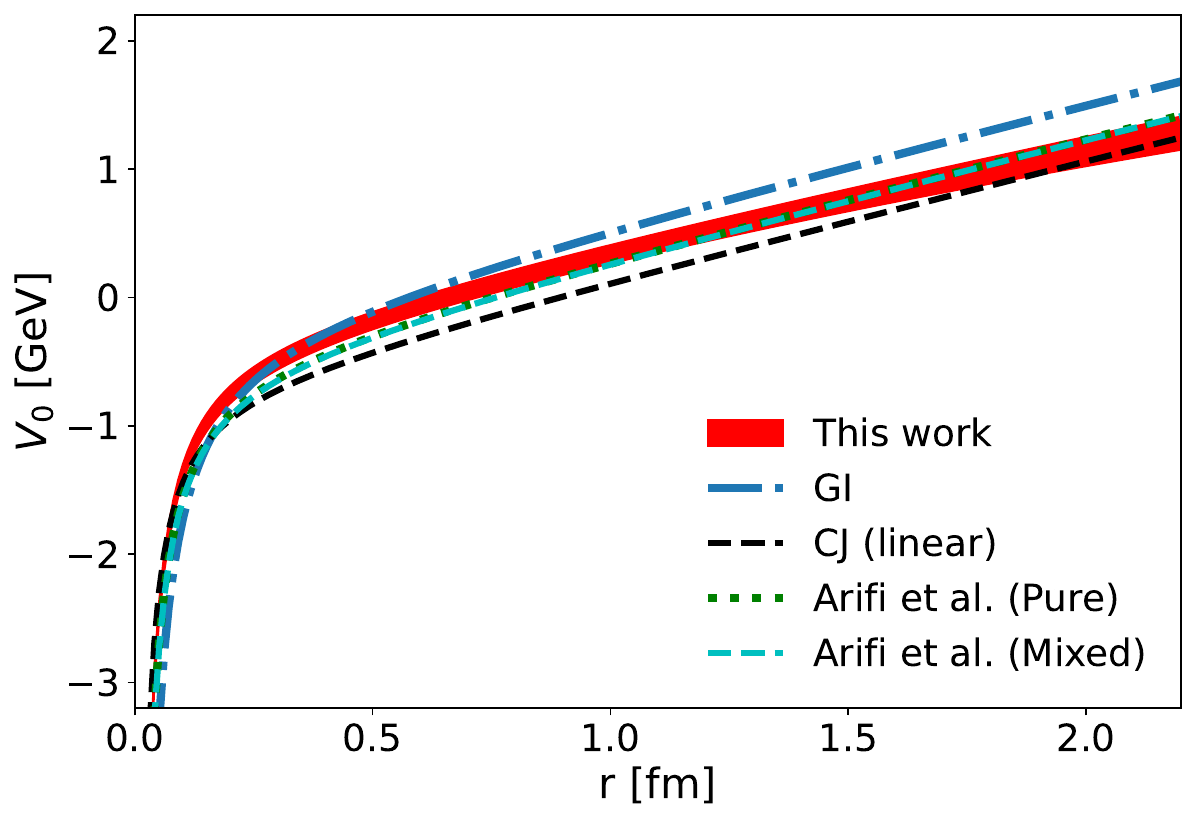}
	\caption{Comparison between our effective potential with Godfrey-Isgur(GI)~\cite{Godfrey:1985xj}, Choi-Ji (CJ) linear~\cite{Choi:2007se}, and Arifi \textit{et al.} models (pure $\theta=0$ and mixed $\theta=12^{\circ}$)~\cite{Arifi:2022pal}. It is shown that our potential is consistent with other models.}
	\label{fig:pot_comparison}
\end{figure}

The obtained model parameters are listed in Table~\ref{tab:parameter} and reasonable agreements between the fitted mass spectra, as well as the decay constants, and the data are obtained as shown in Fig.~\ref{fig:massandDC}. 
We find $\theta_{12} = 12.12^\circ$ and $\theta_{13} = 8.44^\circ$, where $\theta_{12} > \theta_{13}$, a crucial condition for describing the decay constants of the excited states, as will be discussed later. 
Note that the obtained $\beta_{q\bar{q}}$ parameters are common for both pseudoscalar and vector mesons, reflecting the perturbative treatment of the spin-spin interaction.
Figure~\ref{fig:pot_comparison} illustrates the effective potential using the obtained parameters, showing consistency with other models~\cite{Godfrey:1985xj, Arifi:2022pal, Choi:2007se}. However, the potential is suppressed at long distances due to the inclusion of screening confinement effects with $c = 0.027$ GeV, a value similar to those found in Ref.~\cite{Deng:2016ktl}.

\begin{table}[t]
	\begin{ruledtabular}
		\renewcommand{\arraystretch}{1.3}
		\caption{Numerical results of masses and decay constants of the $B_c$ mesons and heavy quarkonia in comparison with experimental and lattice QCD data.
		\label{tab:massandDC}}
		\begin{tabular}{cccccc}
          \multirow{ 2}{*}{States}    & \multicolumn{2}{c}{Mass [GeV]} & \multicolumn{3}{c}{Decay constant [MeV]}  \\
		  & Our & Experiment & Our & Experiment & Lattice\\ \hline  
		 $\eta_c(1S)$   &  2.999(100) &  2.9841(4)      & 349(6) & $335(75)$ & 395(2)\\
   	 $\eta_c(2S)$   &  3.612(97)  &  $3.6377(9)$   & 224(3) & \dots & \dots \\
      $\eta_c(3S)$   &  4.027(96)  &  \dots          & 164(3) & \dots & \dots \\
		 $J/\psi$       &  3.100(96)  &  $3.09690(6)$   & 397(8) & $415(4)$ & 405(6) \\
		 $\psi(2S)$     &  3.678(95)  &  $3.686097(11)$   & 286(5) & $287(3)$ & \dots \\   
		 $\psi(3S)$     &  4.078(94)  &  $4.040(1)$     & 230(4) & $238(5)$ & \dots \\ \hline
		 $B_c(1S)$      &  6.230(84)  &  $6.27447(32)$  & 397(6) & \dots & 427(6) \\
   	 $B_c(2S)$      &  6.845(83)  &  $6.8712(10)$   & 279(4) & \dots & \dots \\    
   	 $B_c(3S)$      &  7.250(82)  &  \dots          & 220(3) & \dots & \dots \\ 
		 $B_c^*(1S)$    &  6.301(83)  &  \dots          & 428(7) & \dots & \dots\\  
   	 $B_c^*(2S)$    &  6.891(82)  &  \dots          & 319(5) & \dots & \dots \\    
   	 $B_c^*(3S)$    &  7.286(82)  &  \dots          & 265(4) & \dots & \dots \\ \hline  
		 $\eta_b(1S)$   &  9.319(101) &  $9.3987(20)$   & 630(9) & \dots & 667(6) \\
   	 $\eta_b(2S)$   &  10.038(99) &  $9.999(4)$     & 457(6) & \dots & \dots \\    
   	 $\eta_b(3S)$   &  10.495(98) &  \dots          & 364(4) & \dots & \dots \\                      
		 $\Upsilon(1S)$ &  9.398(100) &  $9.46040(10)$  & 670(10) & $689(5)$ & 649(31) \\
   	 $\Upsilon(2S)$ &  10.089(98) &  $10.0234(5)$   & 511(8) & $497(5)$  & $481(39)$ \\    
   	 $\Upsilon(3S)$ &  10.535(98) &  $10.3551(5)$   & 428(6) & $430(4)$ & \dots \\ 
        \end{tabular}
		\renewcommand{\arraystretch}{1}
	\end{ruledtabular}
\end{table}

\subsection{Light-front wave functions}

\begin{figure*}[t]
    \centering
    \includegraphics[width=2\columnwidth]{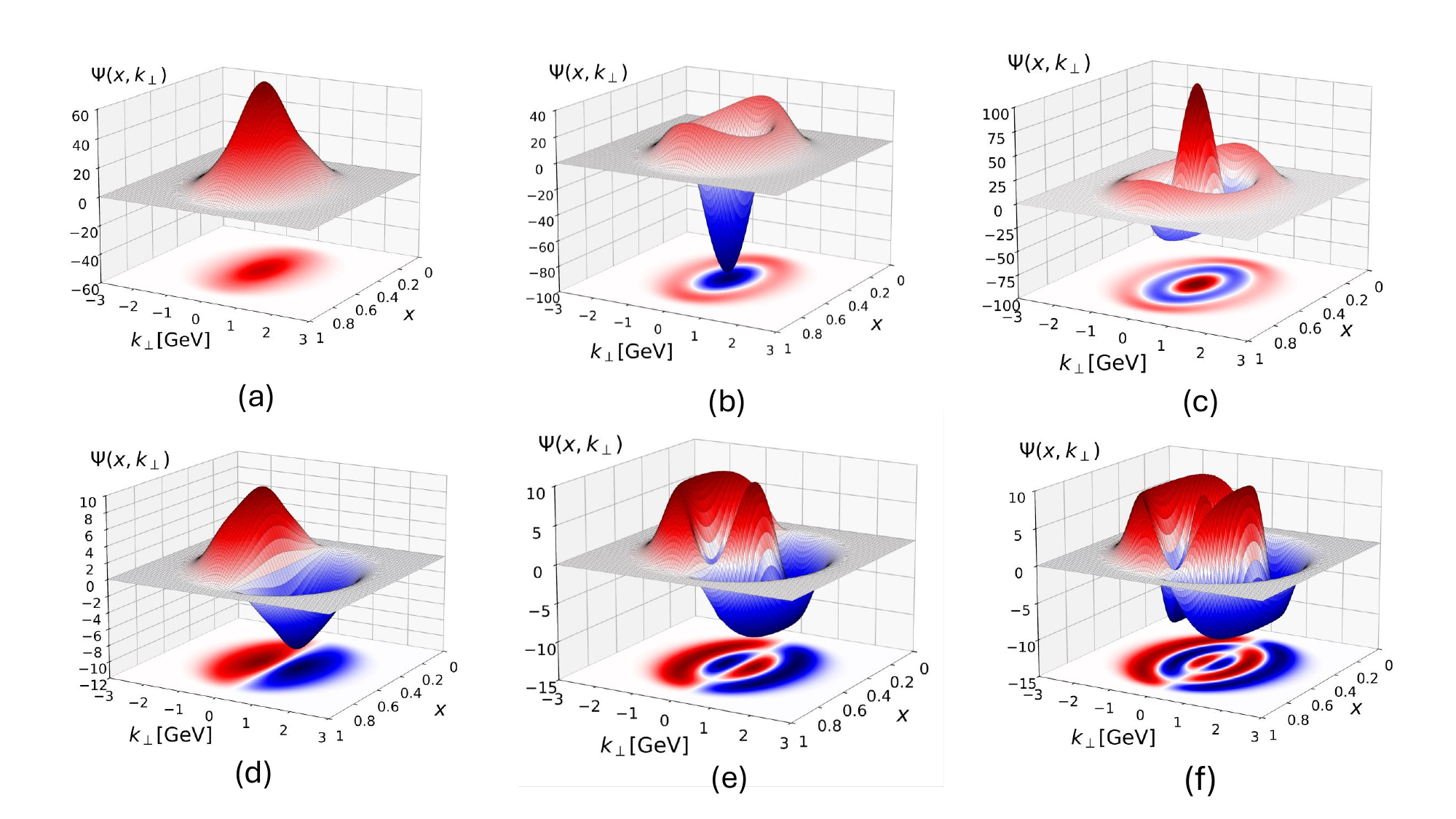}
    \caption{ Upper panels (a)-(c): 3D plots of the ordinary helicity LFWFs $(\Psi_{\uparrow\downarrow} - \Psi_{\downarrow\uparrow})/\sqrt{2}$ for $\eta_c(1S)$, $\eta_c(2S)$, and $\eta_c(3S)$, respectively. Lower panels (d)-(f): 3D plots of the higher helicity LFWFs $\Psi_{\uparrow\uparrow} = \Psi_{\downarrow\downarrow}^*$ for $\eta_c(1S)$, $\eta_c(2S)$, and $\eta_c(3S)$, respectively. Each panel includes a 2D plot of LFWFs at the bottom. The wave function $\Psi(x,k_\perp)$ is shown in units of GeV$^{-1}$ and displays different nodal patterns for each meson.}
    \label{fig:LFWF}
\end{figure*}

The LFWFs obtained for selected mesons, up to the second radial excitation $[\eta_c(1S), \eta_c(2S), \eta_c(3S)]$, are depicted in Fig.~\ref{fig:LFWF}, offering an initial understanding of the meson structures. Alongside the three-dimensional (3D) plots of the LFWFs, two-dimensional (2D) projections are shown at the bottom of each panel. These plots display both the ordinary (upper panel) and higher (lower panel) helicity components of the LFWFs. The nodal structure is also clearly reflected in the ordinary helicity components of the LFWFs, while the higher helicity components oscillate between positive and negative values in the $k_\perp$ domain. Such features are also visualized by BLFQ~\cite{Li:2017uug}. 

The ordinary LFWF components, defined by $[\Psi_{\uparrow\downarrow}(x,\bm{k}_\bot) -\Psi_{\downarrow\uparrow}(x,\bm{k}_\bot)]/\sqrt{2}$, are related to the nonrelativistic wave function. 
In contrast, the higher helicity components, defined by $\Psi_{\uparrow\uparrow} (x,\bm{k}_\bot) = \Psi^*_{\downarrow\downarrow}(x,\bm{k}_\bot)$, are linked to the quark's orbital angular momentum. These higher helicity components have a relativistic origin, arising from the small component of the Dirac spinor, which depends on the orientation of the quantization surface. Their contribution diminishes as the quark mass increases, as implied in the formula for the spin-orbit wave function. 

As shown in Fig.~\ref{fig:LFWF}, the higher helicity components are smaller than the ordinary ones, which is natural since charmonium can be approximated as a nonrelativistic object. Because of the equal mass of the quarks, the peak of the LFWFs is located at $x_{\mathrm{peak}}=0$. However, for $B_c$ mesons with unequal constituent quark masses, the peak will be shifted to $x_{\mathrm{peak}}=m_q/(m_q+m_{\bar{q}})$. 
Compared to that of Ref.~\cite{Li:2017uug}, our 3D plot of LFWFs is roughly consistent after including a $1/2\pi$ factor to match their normalization.
Note that the HO functions usually result in suppressed endpoint behavior near $x=0$ and $x=1$ in the LFWFs. This can be improved by utilizing a more realistic wave function, such as the one provided by the GEM~\cite{Arifi:2024mff}.

\subsection{Mass spectra and splittings}

Our computed mass spectra, along with their comparison to the experimental values from the PDG, are tabulated in Table~\ref{tab:massandDC} and illustrated in the upper panel of Fig.~\ref{fig:massandDC}. The predictions for yet-undiscovered states are also included. The masses were calculated using a screened confining potential, which is known to significantly lower the mass of the excited states compared to those obtained using a linear potential for the charmonia \cite{Ridwan:2024hyh} and also $B_c$ mesons \cite{Hao:2024nqb}. 
Overall, our predictions show good agreement with the experimental data. 
This can be observed through the relative discrepancy, defined as $\Delta M_{\text{err.}} = |(M_{\text{theo.}} - M_{\text{expt.}}) / M_{\text{expt.}} | \times 100 \%$,
where the average discrepancy is less than one percent, specifically around $0.50\%, 0.52\%, 0.85\%$ for $(c\bar{c})$, $(c\bar{b})$, and $(b\bar{b})$, respectively. 
However, the $\Upsilon(3S)$ state has a higher discrepancy, approximately $1.7\%$, compared to the others, which can be attributed to several factors, including the fitting procedure.
In Table~\ref{tab:massandDC}, we also present the mass uncertainties, which range from $0.95\%$ to $3.32\%$, within which most of the experimental data fall.
It is important to note that a more accurate mass spectra prediction could be achieved if we focused solely on fitting the mass spectra. 
However, doing so would result in poorer predictions for the decay constants. Therefore, we have balanced the accuracy to obtain reasonable results for both observables.

Moreover, our model successfully explains the observed hierarchy in the mass spectra as
\begin{eqnarray}
    \Delta M_{\mathcal{P}}^{(n+1)S-nS} &>& \Delta M_{\mathcal{V}}^{(n+1)S-nS}, \label{eq:mass_gap}
\end{eqnarray}
where $\Delta M_{\mathcal{P(V)}}^{(n+1)S-nS} = M_\mathcal{P(V)}^{(n+1)S} - M_\mathcal{P(V)}^{nS}$. 
From Table~\ref{tab:massandDC}, we find that $\Delta M_{\eta_c}^{2S-1S}=612$ MeV $> \Delta M_{\psi}^{2S-1S}=578$ MeV and $\Delta M_{\eta_c}^{3S-2S}=414$ MeV $> \Delta M_{\psi}^{3S-2S}=399$ MeV, which is consistent with Eq.~\eqref{eq:mass_gap} and supports the validity of our approach. 
It is worth noting that this hierarchy cannot be reproduced using the conventional HO functions~\cite{Arifi:2022pal}. 
The calculations above also demonstrate that the mass splitting decreases as radial excitation $n$ increases:
\begin{eqnarray}
    (M_{2S} - M_{1S}) > (M_{3S} - M_{2S}).
\end{eqnarray}
Additionally, the similarity in the mass gap, regardless of the quark flavor content as shown in Fig.~\ref{fig:massandDC}, is due to the competition between the Coulomb and confinement potentials~\cite{Arifi:2022pal}.

\subsection{Decay constants}

The predicted decay constants, along with a comparison to experimental and lattice QCD results, are presented in Table~\ref{tab:massandDC} and depicted in Fig.~\ref{fig:massandDC}. 
Note that the experimental values of the decay constants are derived from the leptonic partial width $\Gamma_{e^- e^+}$ for vector mesons~\cite{ParticleDataGroup:2024cfk}. The decay constant for $\eta_c(1S)$ is based on CLEO data~\cite{CLEO:2000moj}. For $\Upsilon(1S)$, we use the $\Gamma_{e^- e^+}$ value reported in~\cite{CLEO:1998fxo}. For $\psi(3S)$, where $\Gamma_{e^- e^+}$ ranges between 0.6 and 1.4 keV depending on the fit~\cite{Mo:2010bw}, we adopt the fourth solution, which is close to our fit.
Overall, the predicted values are in reasonable agreement with the experimental results, as indicated by the relative discrepancy $\Delta f_{\text{err.}} = |(f_{\text{theo.}} - f_{\text{expt.}}) / f_{\text{expt.}} | \times 100\%$, with an average discrepancy of approximately $2.7\%$. 
The average discrepancy is slightly smaller than that of the mass spectra, which can be understood by the fact that the decay constant is much smaller in magnitude compared to the mass.
In addition, we present the uncertainty of the decay constants in Table~\ref{tab:massandDC} and find that it ranges from $1.19\%$ to $2.07\%$. 
Since we fit only the experimental data, our results show some discrepancy with the lattice QCD data.

In contrast to the mass spectra $M_{\mathcal{P(V)}}$, the decay constants $f_{\mathcal{P(V)}}$ decrease as the radial excitation $n$ increases, indicating the hierarchy 
\begin{eqnarray}
    f_{1S} > f_{2S} > f_{3S}   
\end{eqnarray}
as clearly shown in Fig.~\ref{fig:massandDC}. 
Furthermore, the ratio of the decay constants
\begin{eqnarray}\label{eq:DC_ratio}
     f_{3S}/f_{2S} > f_{2S}/f_{1S}
\end{eqnarray}
also decreases with increasing $n$. 
For instance, the decay constants $f_{J/\psi}, f_{\psi(2S)}, f_{\psi(3S)}$ are 397 MeV, 286 MeV, and 230 MeV, respectively, with $f_{\psi(2S)}/f_{J/\psi} = 0.72$ and $f_{\psi(3S)}/f_{\psi(2S)} = 0.80$.
The observed hierarchy in the decay constants is well reproduced by our model, thanks to the HO basis expansion, which reduces the short-distance component of the wave function~\cite{Arifi:2022pal}. 
We find that $\theta_{12} = 12.12^\circ > \theta_{13} = 8.44^\circ$, which aligns with the observation in Eq.~\eqref{eq:DC_ratio}. 
This is because a larger mixing angle results in a greater suppression of the decay constants for excited states. Additionally, we provide further evidence that this approach is also applicable to the second radial excitation.

\subsection{$M1$ radiative transition}

\begin{figure}[b]
	\centering
	\includegraphics[width=1\columnwidth]{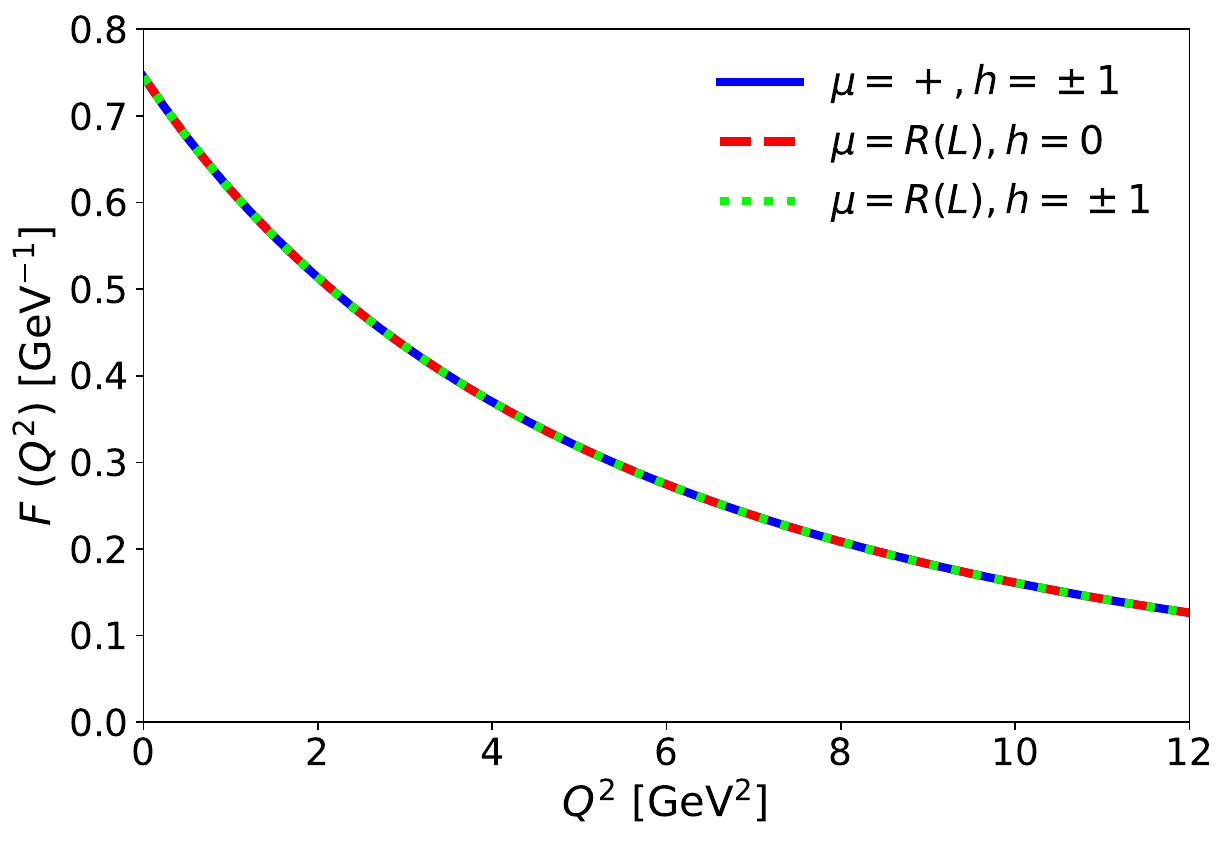}
	\caption{Comparison of transition form factors $F_{\mathcal{VP}}(Q^2)$ of $J/\psi (1S) \rightarrow \eta_c (1S) + \gamma$ computed with various current component $(\mu=+,R(L))$ and $(h=0,\pm1)$ polarizations. It is shown that $F_{\mathcal{VP}}(Q^2)$ is exactly the same regardless the current component and polarization used in the calculation. }
	\label{fig:CompareFF}
\end{figure}

\begin{figure*}[t]
	\centering
	\includegraphics[width=2\columnwidth]{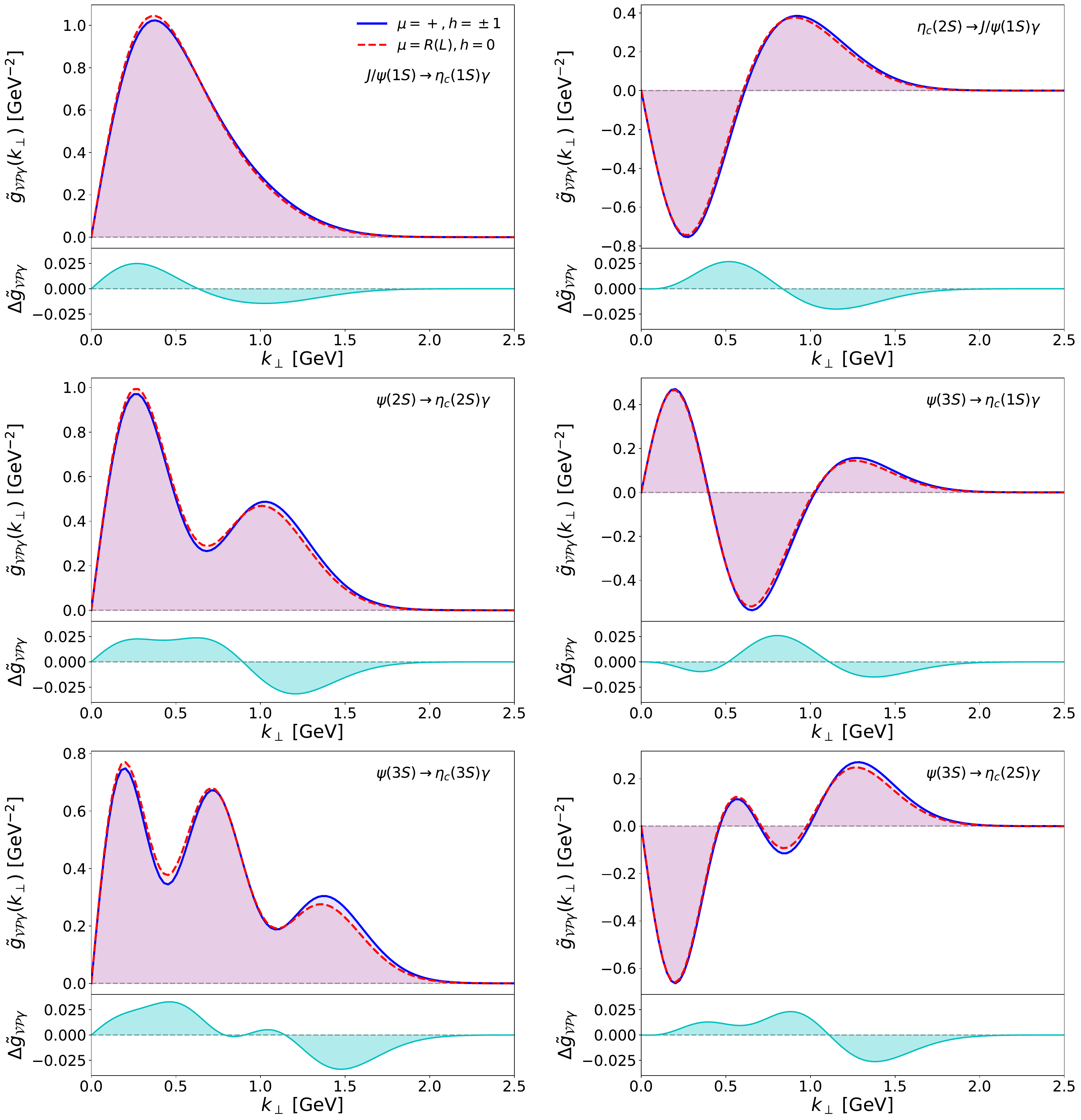}
	\caption{Comparison of the $\tilde{g}_{\mathcal{VP}\gamma}(k_\perp)$ of the $\psi(nS)\to \eta_c(n^\prime S) + \gamma$ transition for $(\mu=+,h=\pm1)$ and $(\mu=R(L),h=0)$ cases. The difference between them $\Delta \tilde{g}_{\mathcal{VP}\gamma} = g^{\mu=+,h=\pm1}_{\mathcal{VP}\gamma} - g^{\mu=R(L),h=0}_{\mathcal{VP}\gamma}$ is shown at the bottom of each panel. Left panels: allowed $(n=n^\prime)$ case and Right panels: hindered $(n\neq n^\prime)$ case.  }
	\label{fig:Integran_cc}
\end{figure*}

Before discussing the numerical results for the $M1$ radiative transition, we examine the self-consistency of the results computed using two different current components: 
(1) the good component ($\mu = +$) with transverse $(h = \pm1)$ polarization~\cite{Choi:2007se}, and 
(2) the transverse component $[\mu = R(L)]$ with longitudinal $(h = 0)$ and transverse $(h = \pm1)$ polarizations. 
To show the self-consistency, let us consider $J/\psi(1S) \rightarrow \eta_c(1S) + \gamma$ for demonstration purposes.
By using the operators given in Table~\ref{tab:operator}, we can obtain the same result for the coupling constant $g_{\mathcal{VP}\gamma}$.
Namely, we confirm that $g_{J/\Psi \eta_c\gamma}^{(\mu=+,h=\pm1)} = g_{J/\Psi \eta_c\gamma}^{(\mu=R(L),h=0,\pm1)}=0.745 $ GeV$^{-1}$ is numerically the same. 

Additionally, we also check that for any $Q^2$ value, with different current components and polarizations, $F_{\mathcal{VP}}(Q^2)$ has the same values, as demonstrated in Fig.~\ref{fig:CompareFF}. 
This is the first time we show the consistency of transition form factors that involve vector mesons, providing more support to the self-consistent LFQM.
Our results are in contrast with those computed in BLFQ, which show different results for different polarizations~\cite{Li:2018uif}.
We argue that the same result can be obtained in the present work since we use the HO basis function, which is inherently rotationally invariant, and the replacement $M\to M_0$ is used when computing the operators in Table~\ref{tab:operator}.

\begin{figure*}[t]
    \centering
    \includegraphics[width=2\columnwidth]{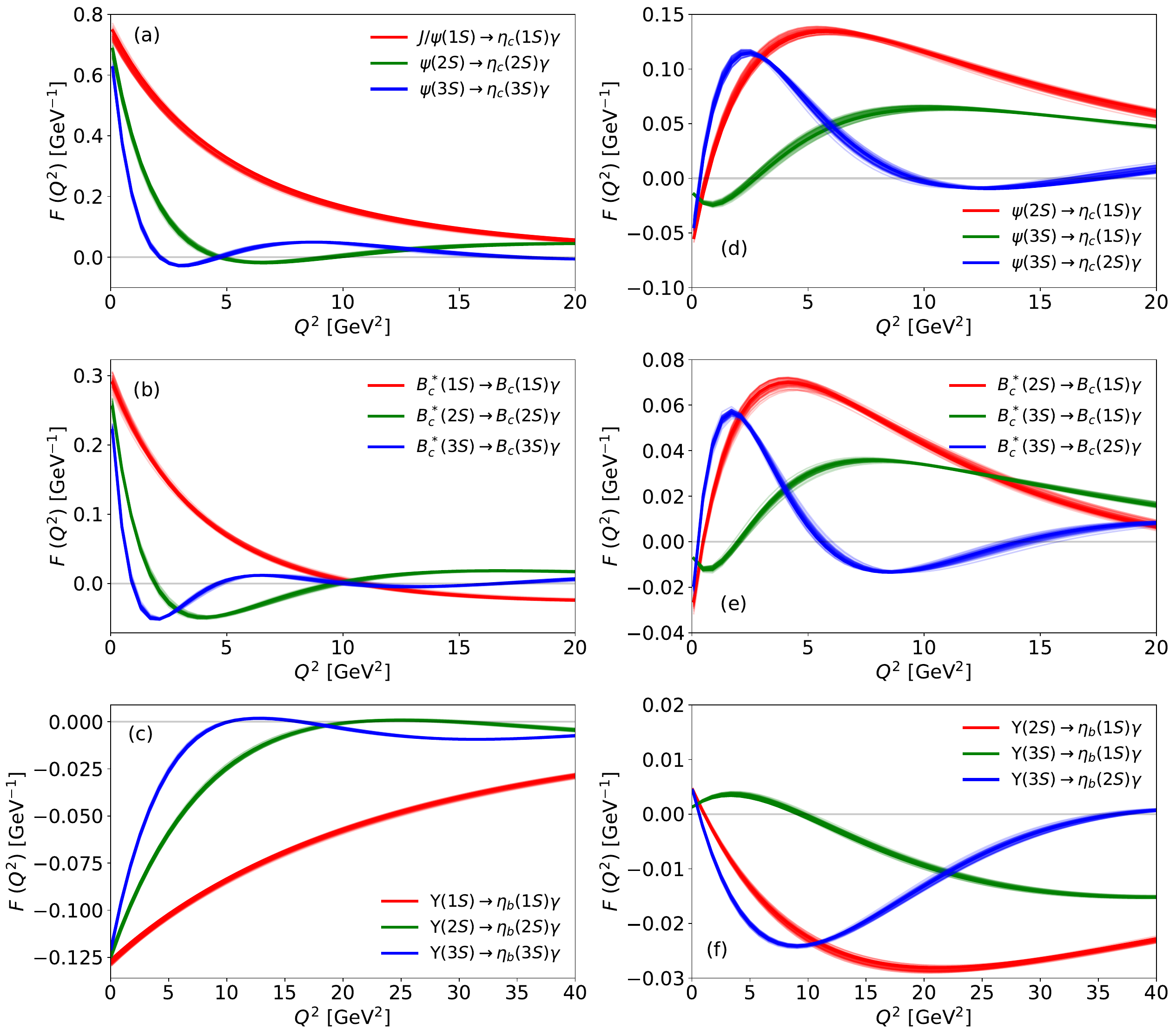}
    \caption{Transition form factors $F_{\mathcal{VP}\gamma}(Q^2)$ for various processes with $\mathcal{V}(nS)\to \mathcal{P}(n^\prime S)+\gamma$; left panels (a-c): allowed $(n=n^\prime)$ case and Right panels (d-f): hindered $(n\neq n^\prime)$ case. Note that we include the numerical uncertainty obtained from the Monte Carlo simulation in the width of curves. We show that $F_{\mathcal{VP}\gamma}(Q^2)$ is much suppressed for the hindered case due to the orthogonality of the wave functions.}
\label{fig:FF}
\end{figure*}

To further explore the difference between the operators with different polarizations,
we can examine the $k_\perp$ dependence of the coupling constant $\tilde{g}_{\mathcal{VP}\gamma}(k_\perp)$.
It is defined by $g_{\mathcal{VP}\gamma} = \int \dd k_\perp\ \tilde{g}_{\mathcal{VP}\gamma}(k_\perp)$ that is associated with the one-loop integral $\tilde{I}_h^\mu(k_\perp)$ computed as
\begin{eqnarray}\label{eq:integrand}
    I^{\mu}_{h} &=& \int_0^\infty \dd k_\perp \tilde{I}_h^\mu(k_\perp) \nonumber\\
    &=& \int_0^\infty \dd k_\perp \biggl[ \int_0^{2\pi}\dd\theta \int_0^1  \dd x \frac{k_\perp}{2(2\pi)^3} \nonumber\\ 
    & & \times \frac{\Phi(x, \bm{k}_\perp^\prime) \Phi(x,\bm{k}_\perp) 
    }{\sqrt{\mathcal{A}^2+\bm{k}_\perp^{\prime 2}}\sqrt{\mathcal{A}^2+\bm{k}_\perp^2}} \mathcal{O}_{h}^{\mu}(x,\bm{k}_\perp) \biggr].\quad 
\end{eqnarray}
The plot of integrands for $\psi(nS)\to \eta_c(n^\prime S) + \gamma$ in two different cases (i) allowed $(n=n^\prime)$ and (ii) hindered $(n\neq n^\prime)$ are displayed in the left and right panels of Fig.~\ref{fig:Integran_cc}, respectively, where each column represents different case with possible combinations of charmonia up to $3S$ state. 
Note that the plot is associated with the integrands (the terms inside the square brackets in Eq.~\eqref{eq:integrand}).
For case (i), the integrand exhibits one peak, two peaks, and three peaks for $1S$, $2S$, and $3S$, respectively, due to the overlap of the initial and final wave functions. This also shows that the overlap is constructive and large, which enhances the coupling constant $g_{\mathcal{VP}\gamma}$. However, for case (ii), the integrands display destructive behavior, primarily due to the orthogonality of the wave functions—for instance, $\phi_{2S}^\mathrm{HO}$ is orthogonal to $\phi_{1S}^\mathrm{HO}$.
In some cases, such as pion emission decay, the orthogonality of the wave function may result in a significant relativistic effect~\cite{Arifi:2021orx}.

Furthermore, it is evident that the $k_\perp$ dependence of the integrands with two different current components for both case (i) and case (ii) are almost the same, as shown in Fig.~\ref{fig:Integran_cc}. Due to their similar shapes, we plot the difference between them, defined by the form $\Delta \tilde{g}_{\mathcal{VP}\gamma} = g^{\mu=+,h=\pm1}_{\mathcal{VP}\gamma} - g^{\mu=R(L),h=0}_{\mathcal{VP}\gamma}$, to see the structure more clearly, as shown in each bottom panel. 
Examining the difference in integrals is crucial, considering that the resulting coupling values should be invariant regardless of the current component $\mu = +$ and $\mu = \perp$ being used. 
It shows that $g_{\mathcal{VP}\gamma}^{\mu = \perp}(k_\perp)$ is typically more enhanced (suppressed) in the low (high) transverse momentum region, although the difference seems to be small, which contrasts with those in BLFQ~\cite{Li:2018uif}. 
Although $\Delta \tilde{g}_{\mathcal{VP}\gamma}$ shows some peaks and dips in the positive and negative regions, it will be zero after taking the $k_\perp$ integration, yielding the self-consistent results as mentioned previously.

The transition form factors $F_{\mathcal{VP}\gamma}(Q^2)$ for various processes $\mathcal{V}(nS)\to \mathcal{P}(n^\prime S)+\gamma$ are presented in Fig.~\ref{fig:FF}, where the allowed ($n=n^\prime$) and hindered ($n\neq n^\prime$) processes are given in the left and right panels, respectively.
One can immediately see that as $Q^2\to 0$ the  $F_{\mathcal{VP}\gamma}(Q^2)$ are significantly enhanced for the allowed case since they are established from an overlap of the wave functions of the same principal and partial wave $nS$. 
Meanwhile, the  $F_{\mathcal{VP}\gamma}(Q^2)$ are much suppressed for the hindered case due to the orthogonality of the initial and final wave functions.
Note that, for the hindered $(n\neq n^\prime)$ case, there is also the $\mathcal{P}(nS)\to \mathcal{V}(n^\prime S)+\gamma$ process, but the  $F_{\mathcal{VP}\gamma}(Q^2)$ is the same with those $\mathcal{V}(nS)\to \mathcal{P}(n^\prime S)+\gamma$. 
This is because we have treated the spin-spin interaction perturbatively.

\begin{table*}[t]
	\centering
 		\renewcommand{\arraystretch}{1.3}
	\caption{Computed coupling constants $g_{\mathcal{VP}\gamma}$ [GeV$^{-1}$] in the present LFQM compared with the experimental data extracted from PDG~\cite{ParticleDataGroup:2024cfk} and results from other models: BLFQ~\cite{Li:2018uif}, GI~\cite{Godfrey:1985xj}, RQM~\cite{Ebert:2002pp}, and NRQM ~\cite{Gao:2024yvz}. } 
	\label{tab:CClist}
	\begin{tabular}{c|c c c c}
		\hline\hline
		Transition & Our & Experiment & \cite{Li:2018uif} & \cite{Godfrey:1985xj}\\
		\hline\hline
		$J/\psi \rightarrow \eta_c(1S) \gamma$    & 0.745(15)    & 0.684(85) & 0.873 & 0.690  \\
		$\psi(2S) \rightarrow \eta_c(2S) \gamma$  & 0.713(14)    & 0.871(313) & 0.739 & 0.680 \\
		$\psi(3S) \rightarrow \eta_c(3S)\gamma$   & 0.688(12)    & \dots & \dots &  \dots  \\
		$\psi(2S) \rightarrow \eta_c(1S) \gamma$  & $-0.0605(37)$ & $-0.040(3)$ & $-0.144$ & $-0.056$  \\
		$\psi(3S) \rightarrow \eta_c(2S) \gamma$  & $-0.0595(33)$ & \dots & \dots &  \dots  \\
		$\psi(3S) \rightarrow \eta_c(1S) \gamma$  & $-0.0108(4)$ & \dots & \dots & \dots \\
		$\eta_c(2S) \rightarrow J/\psi \gamma$    & $-0.0605(37)$ & \dots  & $-0.033(2)$ & \dots  \\
		$\eta_c(3S) \rightarrow \psi(2S) \gamma$  & $-0.0595(33)$ & \dots & \dots & \dots \\
		$\eta_c(3S) \rightarrow J/\psi \gamma$    & $-0.0108(4)$ & \dots  & \dots & \dots \\ \hline 
		$\Upsilon(1S) \rightarrow \eta_b(1S) \gamma$ & $-0.1279(9)$ & \dots & $-0.141$ & $-0.130$  \\
		$\Upsilon(2S) \rightarrow \eta_b(2S) \gamma$ & $-0.1251(9)$ & \dots  & $-0.134$ & $-0.120$ \\
		$\Upsilon(3S) \rightarrow \eta_b(3S) \gamma$ & $-0.1227(8)$ & \dots & $-0.131$ & $-0.120$  \\
		$\Upsilon(2S) \rightarrow \eta_b(1S) \gamma$ & 0.0049(1) & 0.0057(6) & 0.011 & 0.007  \\
            $\Upsilon(3S) \rightarrow \eta_b(2S) \gamma$ & 0.0052(1) & $<$0.0099(8) & 0.009 & 0.007 \\
		$\Upsilon(3S) \rightarrow \eta_b(1S) \gamma$ & 0.00128(3) & 0.0026(1) & 0.005 & 0.004  \\
		$\eta_b(2S) \rightarrow \Upsilon(1S) \gamma$ & 0.0049(1) & \dots & 0.005 & \dots  \\
		$\eta_b(3S) \rightarrow \Upsilon(2S) \gamma$ & 0.0052(1) & \dots & 0.004 & \dots \\	
		$\eta_b(3S) \rightarrow \Upsilon(1S) \gamma$ & 0.00128(3) & \dots & 0.002 & \dots \\ \hline
  	  Transition & Our & \cite{Godfrey:1985xj} & \cite{Ebert:2002pp} & \cite{Gao:2024yvz} \\ \hline
            $B_c^{*+}(1S) \rightarrow B_c^+(1S) \gamma$  & 0.2985(72) & 0.331 & 0.239 & 0.399 \\
		$B_c^{*+}(2S) \rightarrow B_c^+(2S) \gamma$  & 0.2813(65) & 0.354 & 0.268 & 0.178 \\
		$B_c^{*+}(3S) \rightarrow B_c^+(3S)\gamma$   & 0.2677(59) & \dots & \dots & \dots \\
		$B_c^{*+}(2S) \rightarrow B_c^+(1S) \gamma$  & $-0.0340(19)$ & $-0.034$ & $-0.030$ & $-0.032$ \\
		$B_c^{*+}(3S) \rightarrow B_c^+(2S) \gamma$  & $-0.0326(17)$ & \dots &\dots & \dots  \\
		$B_c^{*+}(3S) \rightarrow B_c^+(1S) \gamma$  & $-0.0049(1)$ & \dots &\dots & \dots  \\
		$B_c^+(2S) \rightarrow B_c^{*+}(1S) \gamma$ & $-0.0340(21)$ & $-0.030$ & $-0.042$ & $-0.020$ \\
		$B_c^+(3S)\rightarrow B_c^{*+}(2S) \gamma$  & $-0.0326(17)$ &\dots & \dots & \dots \\
		$B_c^+(3S) \rightarrow B_c^{*+}(1S) \gamma$ & $-0.0049(1)$ &\dots & \dots & \dots \\ 
		\hline\hline
	\end{tabular}
 		\renewcommand{\arraystretch}{1}
\end{table*}

In the left panels of Fig.~\ref{fig:FF} for the allowed case, we observe that  $F_{\mathcal{VP}\gamma}(Q^2)$ for (a) charmonia and (b) $B_c$ mesons transitions between $1S$ states have similar trends where the  $F_{\mathcal{VP}\gamma}(Q^2)$ have the positive value when $Q^2=0$ and tend to decrease with increasing $Q^2$. 
Interestingly,  $F_{\mathcal{VP}\gamma}(Q^2)$ between $2S$ or $3S$ states also display a decreasing behavior as $Q^2$ increases but with some oscillations characterized by the overlap of the initial and final wave functions.
Meanwhile, for bottomonium transitions (c), the  $F_{\mathcal{VP}\gamma}(Q^2)$ has negative values at $Q^2 = 0$, consistent with those in the previous work by Choi~\cite{Choi:2007se}, 
and increases with increasing $Q^2$. Furthermore, the  $F_{\mathcal{VP}\gamma}(Q^2)$ is more extended to the larger value of $Q^2$.

In the right panels of Fig.~\ref{fig:FF} for the hindered case, we can see that the  $F_{\mathcal{VP}\gamma}(Q^2)$ have a small and opposite values at $Q^2=0$ as compared to the allowed case due to the orthogonality of the wave function that we mentioned previously. 
The  $F_{\mathcal{VP}\gamma}(Q^2)$ for $\mathcal{V}(2S) \to \mathcal{P}(1S)$ transitions for (d) charmonia and (e) $B_c$ mesons seem to increase in the small $Q^2$ region before decreasing in the higher $Q^2$ region, whereas the behavior of  $F_{\mathcal{VP}\gamma}(Q^2)$ is opposite for (f) bottomonia transition. 
One can also see that  $F_{\mathcal{VP}\gamma}(Q^2)$ for $\mathcal{V}(3S) \to \mathcal{P}(1S)$ and $\mathcal{V}(3S) \to \mathcal{P}(2S)$ transitions have different behaviors with some more oscillations due to the structure of the overlap of wave function.

\begin{table*}[t]
	\centering
        \renewcommand{\arraystretch}{1.3}
	\caption{Partial decay widths $\Gamma$ [KeV] for various $M1$ radiative decay channel $\mathcal{V}\rightarrow \mathcal{P}\gamma$($\mathcal{P}\rightarrow \mathcal{V}\gamma$) for charmonia and bottomonia. A comparison with experimental data and other models are provided. Note that the $^{\dagger}$ symbol indicates the use of the predicted mass for yet unobserved states.} 
	\label{tab:DWlist}
	\begin{tabular}{c|c c c c}
		\hline
		\hline
		Transition & $\Gamma(\mathrm{Our})$ & Experiment~\cite{ParticleDataGroup:2024cfk} & NRQM~\cite{Soni:2017wvy} & RQM~\cite{Ebert:2002pp} \\
		\hline\hline
		$J/\psi \rightarrow \eta_c(1S) \gamma$ & 1.84(8) & 1.57(37) & 2.72 & 1.05 \\
		$\psi(2S) \rightarrow \eta_c(2S) \gamma$  & 0.14(5) & 0.21(15) & 1.17 & 0.99 \\
		$\psi(3S) \rightarrow \eta_c(3S)^{\dagger} \gamma$ & 2.5(1)$\times 10^{-3}$ & \dots & 9.93 & \dots  \\
		$\psi(2S) \rightarrow \eta_c(1S) \gamma$  & 2.29(27) & 0.99(15) & 7.51 & 0.95 \\
        $\psi(3S) \rightarrow \eta_c(2S) \gamma$  & 0.48(5) & \dots & \dots & \dots \\
		$\psi(3S) \rightarrow \eta_c(1S)\gamma$  & 0.22(1) & \dots & \dots & \dots \\
		$\eta_c(2S) \rightarrow J/\psi \gamma$  & 3.36(41) & $<$1.94(26) $\times 10^{-2}$ & \dots & 1.12 \\ 
		$\eta_c(3S)^{\dagger} \rightarrow \psi(2S) \gamma$  & 0.9(1) & \dots & \dots & \dots \\
		$\eta_c(3S)^{\dagger} \rightarrow J/\psi \gamma$  & 0.48(3) & \dots & \dots & \dots \\ \hline
		$\Upsilon(1S) \rightarrow \eta_b(1S) \gamma$  & 9.29(91)$\times 10^{-3}$ & \dots & 3.77$\times 10^{-2}$ & 5.8$\times 10^{-3}$ \\
		$\Upsilon(2S) \rightarrow \eta_b(2S) \gamma$  & 5.70(28)$\times 10^{-4}$ & \dots & 5.62$\times 10^{-3}$ & 1.4$\times 10^{-3}$ \\
		$\Upsilon(3S)^{\dagger} \rightarrow \eta_b(3S)^{\dagger} \gamma$  & 2.33(40)$\times 10^{-3}$ & \dots & 2.85$\times 10^{-3}$ & 8.0$\times 10^{-4}$ \\
		$\Upsilon(2S) \rightarrow \eta_b(1S) \gamma$  & 1.30(7)$\times 10^{-2}$ & 1.76(38)$\times 10^{-2}$ & 7.72$\times10^{-2}$ & 6.4$\times10^{-3}$ \\
        $\Upsilon(3S) \rightarrow \eta_b(2S) \gamma$  & 2.78(17)$\times 10^{-3}$ & $<$1.26(12)$\times 10^{-2}$ & 3.62$\times 10^{-2}$ & 1.5$\times 10^{-3}$ \\
	  $\Upsilon(3S) \rightarrow \eta_b(1S) \gamma$  & 3.03(14)$\times 10^{-3}$ &  1.03(17)$\times 10^{-2}$ & 7.70$\times 10^{-2}$ & 1.05$\times 10^{-2}$ \\
		$\eta_b(2S) \rightarrow \Upsilon(1S) \gamma$  & 2.53(15)$\times 10^{-2}$ & \dots & \dots & 1.18$\times 10^{-2}$ \\
		$\eta_b(3S)^{\dagger} \rightarrow \Upsilon(2S) \gamma$  & 1.91(10)$\times 10^{-2}$ & \dots & \dots & 2.8$\times 10^{-3}$ \\
		$\eta_b(3S)^{\dagger} \rightarrow \Upsilon(1S) \gamma$  & 1.14(5)$\times 10^{-2}$ & \dots & \dots & 2.4$\times 10^{-2}$ \\
        \hline \hline
	\end{tabular}
 	\renewcommand{\arraystretch}{1}
\end{table*}

\begin{table*}[t]
	\centering
 		\renewcommand{\arraystretch}{1.3}
	\caption{Branching ratio (Br) for various $M1$ radiative decay channel $\mathcal{V}\rightarrow \mathcal{P}\gamma$($\mathcal{P}\rightarrow \mathcal{V}\gamma$) for charmonia and bottomonia. A comparison with experimental data and other models are provided. Note that the $^{\dagger}$ symbol indicates the use of the predicted mass for yet unobserved states. } 
	\label{tab:Brlist}
	\begin{tabular}{c | c c c c}
		\hline
		\hline
		Transition & $\mathrm{Br}(\mathrm{Our})$ & Experiment~\cite{ParticleDataGroup:2024cfk} & NRQM~\cite{Soni:2017wvy} & RQM~\cite{Ebert:2002pp} \\
		\hline\hline
		$J/\psi \rightarrow \eta_c(1S) \gamma$ & 1.98(10)$\times 10^{-2}$& 1.7(0.4)$\times 10^{-2}$ & 2.94$\times 10^{-2}$  & 1.13$\times 10^{-2}$\\
		$\psi(2S) \rightarrow \eta_c(2S) \gamma$  & 4.68(35)$\times 10^{-4}$& 7(5)$\times 10^{-4}$ & 3.98$\times 10^{-3}$ & 3.37$\times 10^{-3}$\\
		$\psi(3S) \rightarrow \eta_c(3S)^{\dagger} \gamma$ & 3.04(48)$\times 10^{-8}$& \dots & 1.24$\times 10^{-4}$ & \dots \\
		$\psi(2S) \rightarrow \eta_c(1S) \gamma$  & 7.81(98)$\times 10^{-3}$& 3.4(0.5)$\times 10^{-3}$ & 2.55$\times 10^{-2}$ & 3.23$\times 10^{-3}$\\
            $\psi(3S) \rightarrow \eta_c(2S) \gamma$  & 5.9(1.1)$\times 10^{-6}$& \dots & \dots & \dots \\
		$\psi(3S) \rightarrow \eta_c(1S)\gamma$  & 2.68(45)$\times 10^{-6}$& \dots & \dots & \dots \\
		$\eta_c(2S) \rightarrow J/\psi \gamma$  & 2.90(55)$\times 10^{-4}$& $<$1.64$\times10^{-6}$ & \dots & \dots\\ 
		$\eta_c(3S)^{\dagger} \rightarrow \psi(2S) \gamma$  & \dots & \dots & \dots & \dots \\
		$\eta_c(3S)^{\dagger} \rightarrow J/\psi \gamma$ & \dots & \dots & \dots & \dots \\ \hline
		$\Upsilon(1S) \rightarrow \eta_b(1S) \gamma$ & 1.72(17)$\times 10^{-4}$ & \dots & 6.98$\times 10^{-4}$ & 1.07$\times 10^{-4}$\\
		$\Upsilon(2S) \rightarrow \eta_b(2S) \gamma$  & 1.80(90)$\times 10^{-5}$& \dots & 1.76$\times 10^{-4}$ & 4.38$\times 10^{-5}$\\
		$\Upsilon(3S)^{\dagger} \rightarrow \eta_b(3S)^{\dagger} \gamma$ & 1.15(11)$\times 10^{-4}$& \dots & 1.40$\times 10^{-4}$ & 3.94$\times 10^{-5}$\\
		$\Upsilon(2S) \rightarrow \eta_b(1S) \gamma$  & 4.09(41)$\times 10^{-4}$ & 5.5$^{+1.1}_{-0.9}$ $\times 10^{-4}$ & 2.25$\times 10^{-3}$ & 2.00$\times 10^{-4}$\\
            $\Upsilon(3S) \rightarrow \eta_b(2S) \gamma$  & 1.38(15)$\times 10^{-4}$& $<$6.7$\times 10^{-4}$ & 1.78$\times 10^{-3}$ & 7.38$\times 10^{-5}$ \\
	    $\Upsilon(3S) \rightarrow \eta_b(1S) \gamma$  & 1.50(16)$\times 10^{-4}$& 5.1(0.7)$\times 10^{-4}$ & 3.79$\times 10^{-3}$ & 5.17$\times 10^{-4}$ \\
		$\eta_b(2S) \rightarrow \Upsilon(1S) \gamma$  & 1.06(1)$\times10^{-6}$& \dots & \dots &  4.91$\times 10^{-7}$\\
		$\eta_b(3S)^{\dagger} \rightarrow \Upsilon(2S) \gamma$  & \dots & \dots & \dots & \dots \\
		$\eta_b(3S)^{\dagger} \rightarrow \Upsilon(1S) \gamma$  & \dots & \dots & \dots & \dots \\
        \hline \hline
	\end{tabular}
 	\renewcommand{\arraystretch}{1}
\end{table*}

\begin{table}[t]
	\centering
 	\renewcommand{\arraystretch}{1.3}
	\caption{Similar with Table~\ref{tab:Brlist}, but for the radiative transitions between $B_c^{(*)}$ states. Note that the results are sensitive to the phase space being used.} 
	\label{tab:DWlist_bc}
	\begin{tabular}{l|c c c }
		\hline
		\hline
		Transition & $\Gamma$(Our) & NRQM~\cite{Gao:2024yvz} & RQM~\cite{Ebert:2002pp}  \\
		\hline\hline
		$B_c^*(1S)^{\dagger} \rightarrow B_c(1S)\gamma$            & 4.03(25)$\times 10^{-3}$ & 4.04 $\times 10^{-2}$ & 3.30 $\times 10^{-2}$ \\
		$B_c^*(2S)^{\dagger} \rightarrow B_c(2S)\gamma$            & 1.50(24)$\times 10^{-3}$ & 3.30 $\times 10^{-3}$ & 1.70 $\times 10^{-2}$ \\
		$B_c^*(3S)^{\dagger} \rightarrow B_c(3S)^{\dagger} \gamma$ & 8.08(37)$\times 10^{-3}$ & \dots & \dots \\
		$B_c^*(2S)^{\dagger} \rightarrow B_c(1S) \gamma$           & 0.57(7) & 0.56 & 0.43 \\
        $B_c^*(3S)^{\dagger} \rightarrow B_c(2S) \gamma$           & 0.17(2) & \dots & \dots \\
		$B_c^*(3S)^{\dagger} \rightarrow B_c(1S) \gamma$           & 4.93(28)$\times 10^{-2}$ & \dots & \dots \\
		$B_c (2S) \rightarrow B_c^*(1S)^{\dagger} \gamma$          & 1.38(16) & 0.14 & 0.49 \\
		$B_c (3S)^{\dagger} \rightarrow B_c^*(2S)^{\dagger} \gamma$& 0.33(3) & \dots & \dots \\
		$B_c (3S)^{\dagger} \rightarrow B_c^*(1S)^{\dagger} \gamma$& 0.12(1) & \dots & \dots \\ 
        \hline \hline
	\end{tabular}
 		\renewcommand{\arraystretch}{1}
\end{table}

The results of the coupling constants $g_{\mathcal{VP}\gamma}$ [GeV$^{-1}$] for the radiative transitions for heavy quarkonia and $B_c$ mesons are listed in Table~\ref{tab:CClist}. 
The coupling constant $g_{\mathcal{VP}\gamma}$ can be converted from the $V(0)$ from the notation in BLFQ~\cite{Li:2018uif} or lattice, by using $g_{\mathcal{VP}\gamma} = 4e_{Q}V(0)/(M_\mathcal{V} + M_\mathcal{P})$.
We observe that the coupling constant values for the $g_{\Upsilon\eta_b\gamma}=-0.1279$ GeV$^{-1}$ is consistent with $-0.124(-0.119)$ GeV$^{-1}$ for linear (HO) potential in the previous LFQM~\cite{Choi:2007se}.
Meanwhile, the $g_{J\Psi\eta_c\gamma}=0.745$ GeV$^{-1}$ is slightly larger than in the previous LFQM with the values 0.681(0.673) GeV$^{-1}$~\cite{Choi:2007se}. 
We observe that the coupling constants for the hindered cases have the opposite sign compared to those of the allowed cases. This can be understood from the orthogonality of the wave functions, which results in a negative sign because the area in the negative region is larger, as illustrated in Fig.~\ref{fig:Integran_cc}.
Note that the sign of $g_{\mathcal{VP}\gamma}$ from the experiment and other calculations is chosen to be consistent with our computed $g_{\mathcal{VP}\gamma}$, as the decay width is the squared of the coupling constant $g_{\mathcal{VP}\gamma}$, as shown in Eq.~\eqref{eq:DWtheo}.
Due to some perturbative treatment of the spin-spin interaction, some hindered transitions have the same values.
For example, in the transition from $2S$ to $1S$ in charmonium, the value obtained is the same, namely $-0.0605$ GeV$^{-1}$.
In Table~\ref{tab:CClist}, we show that our predictions reasonably agree with the results extracted from experiment~\cite{ParticleDataGroup:2024cfk} as well as BLFQ~\cite{Li:2018uif} and GI model~\cite{Godfrey:1985xj}.

We find that the coupling constants $|g_{\mathcal{VP}\gamma}|$ for the allowed case is much larger than those of the hindered case.
It is also evident that the coupling constants follows hierarchy 
\begin{eqnarray}
|g_{J/\psi\eta_c\gamma}| > |g_{B_c^*B_c\gamma}| > |g_{\Upsilon\eta_b\gamma}|    
\end{eqnarray} 
for any transitions. 
Additionally, we observe that 
\begin{eqnarray}
    |g_{J/\psi\eta_c(1S)\gamma}|>|g_{\psi(2S)\eta_c(2S)\gamma}|>|g_{\psi(3S)\eta_c(3S)\gamma}|
\end{eqnarray} 
and it applies to the $B_c$ and bottomonia cases as well.
Such hierarchy can be observed by the other model calculations~\cite{Li:2018uif,Godfrey:1985xj}, 
but the experimental data for the charmonia transitions, namely $g_{J/\psi\eta_c(1S)\gamma}=0.670\ \mathrm{GeV}^{-1} < g_{\psi(2S)\eta_c(2S)\gamma}=0.884$ GeV$^{-1}$, show opposite behavior. 
More precise measurement is useful to clarify this hierarchy.

Once these coupling constants are obtained, the partial decay width $(\Gamma)$ and the branching ratio (Br) can be calculated by taking into account the relevant phase space.
Note that we use $\Gamma_\mathrm{Total}$ from PDG~\cite{ParticleDataGroup:2024cfk} when computing Br.
This will allow us for a direct comparison with the actual experimental data.
In Tables~\ref{tab:DWlist}, we show that our model calculation is consistent with the experimental observations within the same order of magnitude~\cite{ParticleDataGroup:2024cfk}. 
For instance, our calculated value for $\Gamma_{J/\psi\eta_c(1S)\gamma}$ is $1.84\ \text{keV}$, which is consistent with the experimental measurement of $1.57\ \text{keV}$ and aligns with predictions from NRQM and RQM, which are $2.72\ \text{keV}$ and $1.05\ \text{keV}$, respectively.
For the hindered case, our result for $\Gamma_{\psi(2S)\eta_c(1S)\gamma}$ is $2.29\ \text{keV}$, which agrees reasonably with the experimental data of $0.99\ \text{keV}$ and RQM prediction of $0.95\ \text{keV}$. However, it is notably smaller compared to the NRQM result of $7.51\ \text{keV}$.
For $\Gamma_{\psi(3S)\eta_c(3S)^\dagger\gamma}$, our prediction of $1.98 \times 10^{-3}$ keV is much smaller than the NRQM value of 9.93 keV reported by Soni et al.~\cite{Soni:2017wvy}. However, it is relatively consistent with the value of 0.088 keV in the NRQM study by Deng et al.~\cite{Deng:2016stx}.
The agreements with experimental data and other models are also obtained for the $M1$ transitions between bottomonia.
Furthermore, the corresponding branching ratios roughly agree in the same order of magnitude with experimental data.
However, we note that for some cases, our results are rather different from the NRQM as shown in Table~\ref{tab:Brlist}.

Although the coupling constant for hindered case is small $g=-0.0605$ GeV$^{-1}$, the width can be much enhanced $\Gamma=2.25$ KeV because of the large phase space such as in the $\psi(2S)\to\eta_c(1S)\gamma$ case.
On the other hand, despite the large coupling constant $g=0.6877$ GeV$^{-1}$, the width is suppressed due to the small phase space as for the $\psi(3S)\to\eta_c(3S)\gamma$ case.
Another interesting observation is that the decay width of $\mathcal{P}\rightarrow \mathcal{V}\gamma$  channels have larger $\Gamma$ than  $\mathcal{V}\rightarrow \mathcal{P}\gamma$  channels,  such as in $\eta_c(2S) \rightarrow J/\psi(1S) + \gamma$ with $\Gamma = 3.36$ KeV is larger than $\psi(2S) \rightarrow \eta_c(1S) + \gamma$ with $\Gamma = 2.29$ KeV. 
The difference mainly arises from the phase space, as their couplings are the same, as shown in Table~\ref{tab:CClist}. It is important to experimentally verify this observation, as no data are currently available.

\begin{figure}[t]
	\centering
	\includegraphics[width=1\columnwidth]{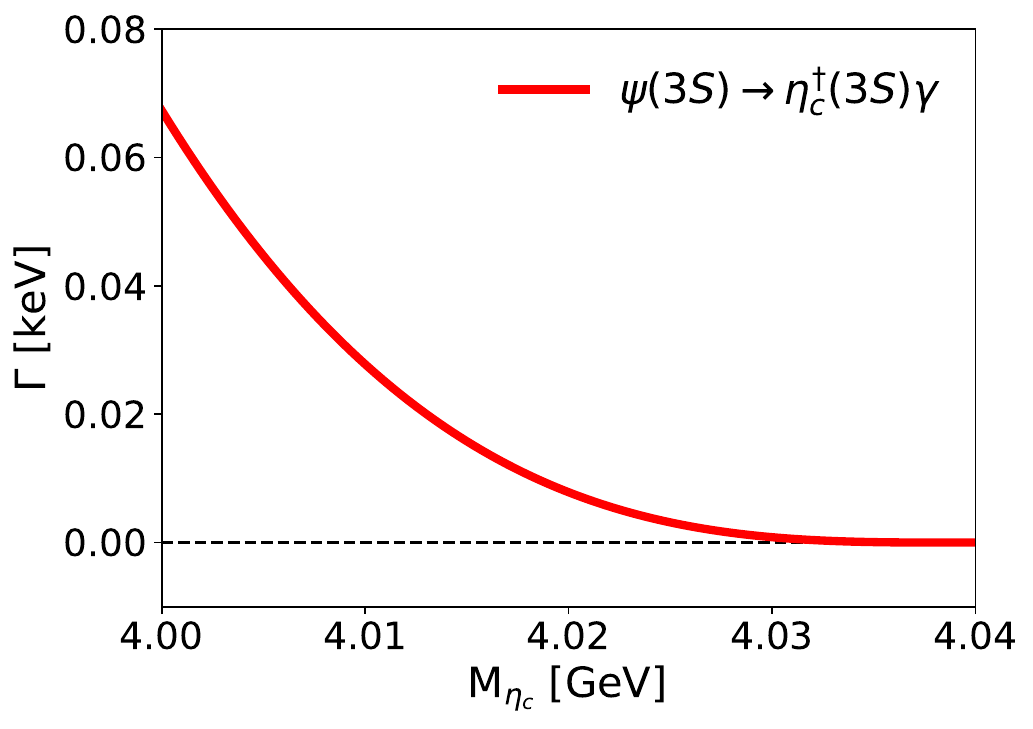}
	\caption{Decay width $\Gamma$ [KeV] of $\psi(3S)\to \eta_c^\dagger(3S)\gamma$ as a function of $\eta_c(3S)$ mass. The width can be up to around 0.06 KeV when the $\eta_c(3S)$ mass is set to be 4 GeV.}
	\label{fig:gamma_mass}
\end{figure}

For the $B_c$ case, since there is no experimental data, we only show our result for the decay width and compared with other model calculations as shown in Table~\ref{tab:DWlist_bc}.
Although some results agree with each other, there are several discrepancies.
However, we note that the results of the $\Gamma$ is sensitive to the phase space being used. 
In this case, the discussion of the coupling constant, as shown in Table~\ref{tab:CClist}, is more relevant when comparing it with other models.

Unlike the coupling constant $g_{\mathcal{VP}\gamma}$, the decay width is influenced by the phase space which depends on the masses of initial and final state mesons.
For some cases where the meson state is not discovered yet, we use the model input for the mass of the involving mesons indicated by the $^{\dagger}$ symbol. 
But, one should note that the prediction has some uncertainties because of the estimated phase space being used.
Figure~\ref{fig:gamma_mass} shows some variations of the decay width as a function of the $\eta_c(3S)$ mass
for the $\psi(3S) \rightarrow \eta_c(3S)^{\dagger} + \gamma$ transition. 
The decay width may vary by up to around 0.06 keV when the $\eta_c(3S)$ mass is set to 4 GeV, which is an order of magnitude larger than the values presented in Table~\ref{tab:DWlist}, which uses the predicted mass from Table~\ref{tab:massandDC}.
Such discussion can be applied to other undiscovered states.
Nevertheless, the predicted values can be improved by reexamining the transitions once the state is found.

\section{Conclusion} 
\label{sec:conclusion}

In this work, we have analyzed the mass spectra and other properties, such as decay constants and $M1$ radiative transitions, of heavy quarkonia and $B_c$ mesons from the ground state up to the second radially excited state by using the LFQM. 
To obtain the LFWFs, we have carried out a variational analysis by employing the QCD-motivated effective potential and utilizing the HO basis expansion up to $3S$ state which is then transformed to LFWFs. 
The model parameters are obtained by fitting the decay constants and mass spectra and found to be in reasonable agreement with the data~\cite{ParticleDataGroup:2024cfk}.

We found that the small mixing that determines the expansion coefficients is crucial for correctly reproducing the decay constant and mass hierarchies. 
Using the obtained LFWFs, we observed that the results for the coupling constants and decay widths of the $M1$ radiative transitions between various states also show reasonable agreement with available experimental data and lattice QCD. 
These findings demonstrate that a simple approach can capture essential features, providing a reasonable explanation for the experimental observations of mass spectra, decay constants, and $M1$ radiative transitions of excited $B_c$ and heavy quarkonia.
However, more experimental data are needed to verify hierarchies for the $M1$ transitions predicted in the present work.

We also pay special attention to the self-consistency of the $M1$ radiative transitions with both longitudinal $(h=0)$ and transverse $(h=\pm1)$ polarizations by considering the good ($\mu=+$) and transverse ($\mu=\perp$) components of currents.
For the first time, we have shown that the coupling constants $g_{\mathcal{VP}\gamma}$ as well as the transition form factors $F_{\mathcal{VP}\gamma}(Q^2)$ with any $Q^2$ are the same, regardless of the current component and the polarizations being used.
This is in contrast to the case of BLFQ~\cite{Li:2018uif}, where different polarizations lead to different results.
We emphasize that the self-consistency can be obtained by replacing the $M\to M_0$ in the matrix elements, which is in accordance with the Bakamjian-Thomas (BT) construction that has been discussed in details in the previous works~\cite{Arifi:2023uqc,Arifi:2022qnd,Choi:2021mni,Choi:2013mda,Choi:2024ptc}.
We have shown that the observables with different polarizations are useful
to quantify the rotational symmetry breaking of the wave function. 
The use of the HO basis functions will naturally satisfy $\expval{k_\perp^2}=\expval{2k_z^2}$.

As an outlook, further investigation is required regarding the wave function model, the fitting procedure, and the potentials employed~\cite{Arifi:2024mff}, as they influence the parameter values needed to determine the observables. For a self-consistent study, it would be interesting to investigate the matrix elements for the minus (bad) component ($\mu = -$) to compare with the present results, although complications such as modifications to the Lorentz structure and nonzero helicity flip contributions may arise~\cite{Choi:2024ptc}. Such a study could alternatively be verified using the Bethe-Salpeter model by applying the replacement $M \to M_0$~\cite{Arifi:2022qnd}. More detailed investigations will be addressed in future work. Furthermore, exploring excited quarkonia and $B_c$ mesons beyond the leading Fock space is crucial to clarifying the underlying dynamics.

\section*{Acknowledgements}

The authors thank Zulkaida Akbar, Makoto Oka, Atsushi Hosaka, Chueng-Ryong Ji, Ho-Meoyng Choi, and Hyun-Chul Kim for some useful discussions. 
M. R. acknowledges the hospitality of the RIKEN Nishina Center during his stay, which facilitated the completion of this work, as well as the stimulating discussions at the RIKEN workshop on the quark structure of hadrons 2024.
A.J.A. is supported by the RIKEN Special Postdoctoral Researcher (SPDR) Program.
T.M. is supported by the PUTI Q1 Grant from University of Indonesia under contract No. NKB-441/UN2.RST/HKP.05.00/2024. 

\appendix

\section{Wave function and mass formula}
\label{app:WF}

\subsection{Wave function}

In this work, we use the HO basis functions and consider up to three basis functions.
The wave function is written as
\begin{eqnarray}
        \Phi_i &=& R_{ij}~\phi^\mathrm{HO}_j,\\
    \begin{pmatrix}
        \Phi_{1S} \\
        \Phi_{2S} \\
        \Phi_{3S}
    \end{pmatrix}
        &=&
    \begin{pmatrix}
    c_1^{1S} & c_2^{1S} & c_3^{1S} \\
    c_1^{2S} & c_2^{2S} & c_3^{2S} \\
    c_1^{3S} & c_2^{3S} & c_3^{3S} \\
    \end{pmatrix}
    \begin{pmatrix}
        \phi_{1S}^\mathrm{HO} \\
        \phi_{2S}^\mathrm{HO} \\
        \phi_{3S}^\mathrm{HO}
    \end{pmatrix},
\end{eqnarray}
where the mixing matrix $\bm{R}$ is explicitly given by
\begin{eqnarray}
\bm{R} = 
\begin{pmatrix}
    1 & 0 & 0 \\
    0 & c_{23} & s_{23} \\
    0 & -s_{23} & c_{23}  \\
\end{pmatrix} 
\begin{pmatrix}
    c_{13} & 0 & s_{13} \\
    0 & 1  & 0 \\
    -s_{13} & 0 & c_{13} \\
\end{pmatrix} 
\begin{pmatrix}
    c_{12} & s_{12} & 0 \\
    -s_{12} & c_{12}  & 0 \\
    0 & 0 & 1 \\
\end{pmatrix}
\nonumber\\
= 
\begin{pmatrix}
    c_{12}c_{13} & s_{12}c_{13} & s_{13} \\
    -s_{12}c_{23} - c_{12}s_{23}s_{13} & c_{12}c_{23} - s_{12}s_{23}s_{13} & s_{23}c_{13} \\
    s_{12}s_{23} - c_{12}c_{23}s_{13} & -c_{12}s_{23} - s_{12}c_{23}s_{13} & c_{23}c_{13} 
\end{pmatrix}, \nonumber\\
\end{eqnarray}
with $c_{ij}(s_{ij}) = \cos\theta_{ij}(\sin\theta_{ij})$ and $i,j=1,2,3$.
For instance, the wave function of $1S$ state meson is
\begin{eqnarray}
    \Phi_{1S} = c_1^{1S} \phi_{1S} + c_2^{1S} \phi_{2S} + c_3^{1S} \phi_{3S},
\end{eqnarray}
with 
\begin{eqnarray}
    c_1^{1S} = c_{12}c_{13}, \quad 
    c_2^{1S} = s_{12}c_{13}, \quad 
    c_3^{1S} = s_{13}.
\end{eqnarray}
The wave functions of the $2S$ and $3S$ states are obtained by replacing the expansion coefficients appropriately as
\begin{eqnarray}
    c_1^{2S} &=& -s_{12}c_{23} - c_{12}s_{23}s_{13},\\
    c_2^{2S} &=& c_{12}c_{23} - s_{12}s_{23}s_{13}, \\
    c_3^{2S} &=& s_{23}c_{13},   
\end{eqnarray}
and 
\begin{eqnarray}
    c_1^{3S} &=& s_{12}s_{23} - c_{12}c_{23}s_{13},\\
    c_2^{3S} &=& -c_{12}s_{23} - s_{12}c_{23}s_{13}, \\
    c_3^{3S} &=& c_{23}c_{13},   
\end{eqnarray}
respectively.
The sum of the squared expansion coefficients is 
\begin{eqnarray}
    c_1^2 + c_2^2 + c_3^2 = 1,
\end{eqnarray}
ensuring that the squared wave function $|\Phi_{nS}|^2$ is normalized to unity.

\subsection{Mass formula}

The mass formula for mesons using the HO basis functions, expanded up to $3S$ basis functions, is given by
\begin{eqnarray}
    M_{q\Bar{q}} &=& \bra{\Psi_{q\Bar{q}}} (H_0 + V_{q\Bar{q}})\ket{\Psi_{q\Bar{q}}} \nonumber \\
    &=&\expval{ H_0} + \expval{V_{\rm Conf}} + \expval{V_{\rm Coul}} + \expval{V_{\rm Hyp}}.\quad 
\end{eqnarray}
The kinetic matrix elements are given by
\begin{eqnarray}
    \expval{H_0} &=& \frac{\beta}{120\sqrt{\pi}}  \sum_{j=1,2} \biggl[  ( g_1 + g_2 z_i^2) z_i {\rm e}^{z_i/2} K_1 \left[ \frac{z_i}{2}\right]  \nonumber\\
    & & + g_3  z_i^2 (z_i^2 -3) {\rm e}^{z_i/2} K_2 \left[ \frac{z_i}{2}\right] \nonumber \\ 
    & & + 15\sqrt{\pi} \biggl( g_4 U(-1/2,-2,z_i) + g_5 U(-1/2,-4,z_i) \nonumber\\
    & &  + g_6 U(-1/2,-5,z_i) \biggr) \biggl],
\end{eqnarray}
where $z_i = m_i^2/\beta^2$, $K_n(x)$ is the modified Bessel function of second kind of order \textit{n}, and $U(a,b,z)$ is the Tricomi's (confluent hypergeometric) function.
The coefficients are given by
\begin{eqnarray}
    g_1 &=& 120 c_1^2 - 120\sqrt{6} c_1 c_2 + 180 c_2^2 + 60 \sqrt{30} c_1 c_3 \nonumber\\
    & & - 180\sqrt{5} c_2 c_3 + 225 c_3^2, \\
    g_2 &=& 40 c_2^2 + 8\sqrt{30} c_1 c_3 -104\sqrt{5} c_2 c_3 +  260 c_3^2, \\
    g_3 &=& -4(10 c_2^2 -26\sqrt{5} c_2 c_3 + 2 \sqrt{30}c_1 c_3 + 65 c_3^2),\quad\quad   
\end{eqnarray}
and 
\begin{eqnarray}
    g_4 &=& 4(-6 c_2^2 + 9\sqrt{5} c_2 c_3 - 15 c_3^2 \nonumber\\
    & & + 2 \sqrt{6}( c_1 c_2 - \sqrt{5} c_1 c_3)),\\
    g_5 &=& 28 (\sqrt{5} c_2 c_3 - 5 c_3^2),\\
    g_6 &=& 63 c_3^2,
\end{eqnarray}
Several types of confining potentials are used in the literature. 
Here, we provide two common ones. The matrix element of the linear confining potential is given by
\begin{eqnarray}
   \expval{V^{\rm Lin}_{\rm Conf}} &=& a  + \frac{b}{\beta \sqrt{\pi}}\biggl(2c_1^2 + 3c_2^2 + \frac{15}{4}c_3^2 \nonumber\\
   & & - 2\sqrt{\frac{2}{3}}c_1 c_2 - \sqrt{\frac{2}{15}}c_1 c_3  - \sqrt{5}c_2 c_3 \biggr),  \quad \quad 
\end{eqnarray}
while the matrix element for the screened potential is given by
\begin{eqnarray}
  \expval{V^{\rm scr}_{\rm Conf}} &=& a + \frac{b}{\mu} + \frac{b}{3840\sqrt{\pi}\beta^{10}\mu} \biggl[ 2\beta \mu \bigl( d_1 \beta^8 \nonumber\\
  & & + d_2 \beta^6\mu^2 + d_3 \beta^4\mu^4 + d_4 \beta^2 \mu^6 + d_5 \mu^8 \bigr) \nonumber\\
  & & - \sqrt{\pi}{\rm e}^{\mu^2/4\beta^2} ( f_1 \beta^{10} + f_2 \beta^8 \mu^2 + f_3 \beta^6 \mu^4 \nonumber\\
  & & + f_4 \beta^4 \mu^6 + f_5 \beta^2 \mu^8 + f_6 \mu^{10} ) {\rm erfc}\left[\frac{\mu}{2\beta}\right]\biggl],\nonumber\\ \quad \quad \quad 
\end{eqnarray}
where 
\begin{eqnarray}
    d_1 &=& 32(60 c_1^2 - 4\sqrt{6}(10 c_1 c_2 + \sqrt{5} c_1 c_3) \nonumber\\
    & & + 15(8c_2^2 -4\sqrt{5}c_2 c_3 + 11 c_3^2)), \\
    d_2 &=& 32(40 c_2^2 -48\sqrt{5}c_2 c_3 + 115 c_3^2 \nonumber\\
    & & + 2\sqrt{6}(-5 c_1 c_2 + 2\sqrt{5} c_1 c_3),  \\
    d_3 &=& 8 (10c_2^2 -28\sqrt{5} c_2 c_3 + 2\sqrt{30} c_1 c_3 + 89 c_3^2  ), \quad\quad  \\
    d_4 &=& -8(\sqrt{5} c_2 c_3 - 6 c_3^2 ),\\
    d_5 &=& c_3^2.
\end{eqnarray}
and 
\begin{eqnarray}
    f_1 &=& 3840,\\
    f_2 &=& 1920( c_1^2 + 3c_2^2 + 5 c_3^2 - \sqrt{6} c_1 c_2 - 2\sqrt{5} c_2 c_3),\quad \quad  \\
    f_3 &=& 160( 9 c_2^2 + 30 c_3^2 + \sqrt{30} c_1 c_3 - 2\sqrt{6}c_1 c_2\nonumber\\
    & &  - 12\sqrt{5} c_2 c_3  ),\\
    f_4 &=& 16( 5 c_2^2 + 50 c_3^2 - 15\sqrt{5} c_2 c_3 + \sqrt{30} c_1 c_3 ), \\
    f_5 &=& 50 c_3^2 - 8\sqrt{5}c_2 c_3, \\
    f_6 &=& c_3^2.
\end{eqnarray}
with 
\begin{eqnarray}
    {\rm erfc}(z) = 1 - {\rm erf}(z).
\end{eqnarray}
Lastly, the matrix elements for the Coulomb and Hyperfine potentials are given by
\begin{eqnarray}
  \expval{V_{\rm Coul}} &=& - \frac{\beta \alpha_s}{45\sqrt{\pi}} \biggl( 120c_1^2 +100c_2^2 + 89 c_3^2 \nonumber\\
  & & + 40\sqrt{6} c_1 c_2 + 12\sqrt{30} c_1 c_3  + 44\sqrt{5}c_2 c_3 \biggr),\nonumber\\
\end{eqnarray}
and
\begin{eqnarray}
  \expval{V_{\rm Hyp}} &=& \frac{\expval{\textbf{S}_q\cdot\textbf{S}_{\Bar{q}}}\beta^3 \alpha_s}{3 m_1 m_2\sqrt{\pi}} \biggl( \frac{32}{3}c_1^2 + 16 c_2^2 + 20 c_3^2 \nonumber\\
  & & + 32\sqrt{\frac{2}{3}} c_1 c_2 + 16\sqrt{\frac{10}{3}} c_1 c_3 + 16 \sqrt{5} c_2 c_3\biggr). \nonumber\\ 
\end{eqnarray}
The mass of the meson for $1S$, $2S$, and $3S$ states is computed by replacing $c_i$ in the mass formula with $c_i^{1S}$, $c_i^{2S}$, or $c_i^{3S}$, respectively.

\section{Fitting procedure and error propagation}
\label{app:fitting}

Here we explain the fitting procedure to determine the model parameters and show how we obtain the error propagation numerically.
This is an attempt to fix the model parameters using a statistical method and quantify the statistical uncertainty of the parameters, 
instead of determining them via the trial-and-error method.
We started with a large number of parameters, but the variational principle put some constraints and reduced them significantly. 
Additionally, we assume that $\theta_{13}=\theta_{23}$ and $b=0.18$ GeV$^2$ to simplify the parameterization.

As mentioned earlier, this work is based on the variational analysis where the variational parameters $\beta$ are constrained by means of 
\begin{eqnarray}\label{eq:variational}
    \frac{\partial \expval{H_0(\beta_{q\bar{q}}) + V_{\mathrm{Conf}}(\beta_{q\bar{q}}) + V_{\mathrm{Coul}}(\beta_{q\bar{q}}) } }{\partial \beta_{q\bar{q}}} = 0, 
\end{eqnarray}
where we treat the hyperfine interaction perturbatively, which works sufficiently well for mesons that contain heavy quarks.
The above equation enables us to express the parameter $\beta$ in term of other parameters for each meson as
\begin{eqnarray}
	\beta_{c\bar{c}} &\to& 	\beta_{c\bar{c}}  (\theta_{12},\theta_{13},m_q,m_{\bar{q}}, a, c, \alpha_s), \\
	\beta_{b\bar{c}} &\to& 	\beta_{b\bar{c}}  (\theta_{12},\theta_{13},m_q,m_{\bar{q}}, a, c, \alpha_s),\\
	\beta_{b\bar{b}} &\to& 	\beta_{b\bar{b}}  (\theta_{12},\theta_{13},m_q,m_{\bar{q}}, a, c, \alpha_s),
\end{eqnarray}
where we assume that the potential and mixing parameters are commons for each meson.

To summarize, we have seven free parameters ($\theta_{12},\theta_{13},m_q,m_{\bar{q}}, a, c, \alpha_s$) to fit.
By using these parameters, $\beta_{q\bar{q}}$ are automatically determined from the constraint given by Eq.~\eqref{eq:variational}.
Then, we fit these parameters to the experimental data of mass spectra and decay constants by using the iMinuit package~\cite{iminuit,James:1975dr}, simultaneously, where the $\chi^2$ is defined by Eq.~\eqref{eq:chi2}. The  parameter uncertainty is also extracted from the Hesse Error.
Note that we have introduced $\sigma_\mathrm{mod}$ in an attempt to reduce the fitting bias due to the different precision. 

\begin{figure}[t]
	\centering
	\includegraphics[width=1\columnwidth]{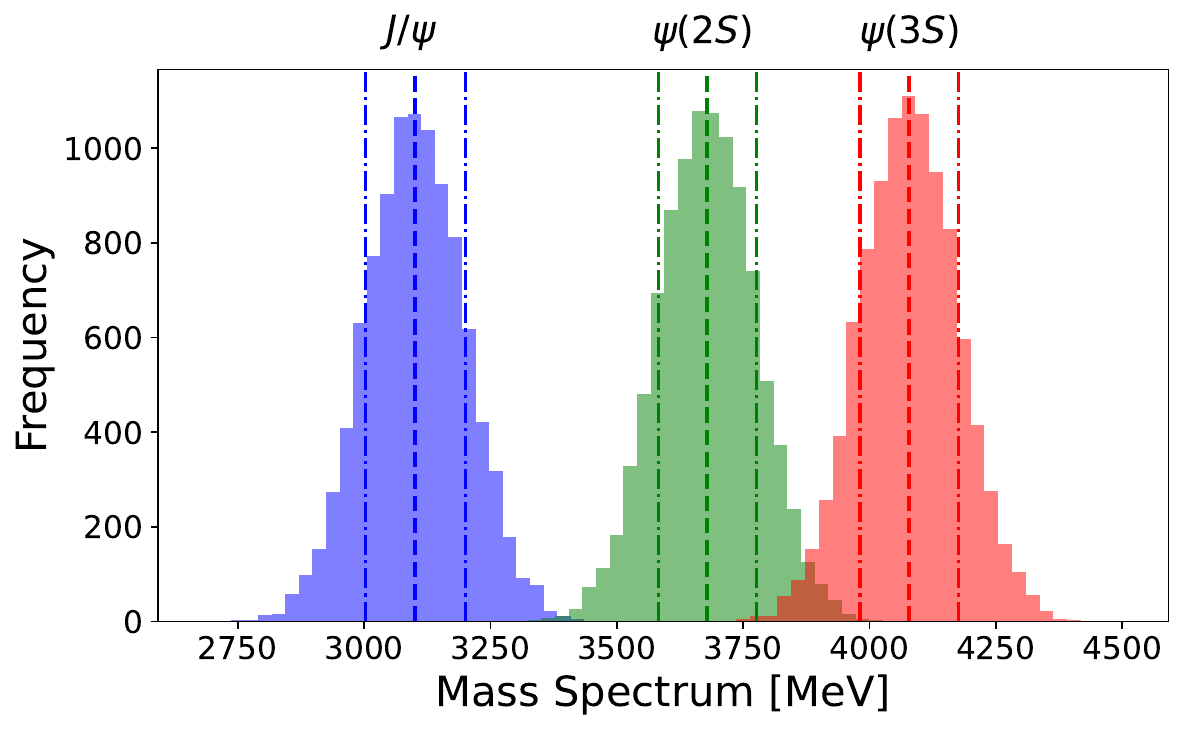}
	\caption{Histogram of the masses of $J/\psi$, $\psi(2S)$, and $\psi(3S)$ computed using the Monte Carlo method to quantify uncertainty. The mean and standard deviation for each mass are represented by the dashed and dash-dotted lines, respectively. \label{fig:histogram} }
\end{figure}

To estimate the uncertainty, we utilize the Monte Carlo method, a statistical approach that assesses the propagation of uncertainties through random sampling to evaluate how the uncertainties of input parameters  affect the output.
In the Monte Carlo approach, each input parameter $x_i$ is associated with a probability distribution characterized by a mean value $\mu_i$ and a standard deviation (uncertainty) $\sigma_i$. 
Parameters are then sampled from these distributions, using the normal distributions
\begin{equation}
    x_i \sim \mathcal{N}(\mu_i, \sigma_i^2).
\end{equation}
For each set of sampled parameters, the output $y$ is calculated, resulting in $y^k = f(x_1^k, x_2^k, \ldots, x_n^k)$ for $k = 1, 2, \ldots, N$. 
The output is characterized by its mean value $\bar{y}$, which represents the expected value, and the standard deviation $\sigma_y$, which measures the spread of the output values around the mean.
As for a rough estimation of the uncertainty, we do not consider the covariance matrix in this work.
For example, we present the histogram of the masses of $J/\psi$, $\psi(2S)$, and $\psi(3S)$ in Fig.~\ref{fig:histogram}. It shows that the mass distributions approximately follow a Gaussian distribution, allowing us to extract the mean and standard deviation.

\section{Some conventions and definitions}
\label{app:definitions}

In this work, we use the Lepage-Brodsky convention~\cite{Lepage:1980fj}, given by
\begin{eqnarray}
\begin{pmatrix}
x^+\\
x^-
\end{pmatrix}
=
\begin{pmatrix}
1 & 1 \\
1 & -1
\end{pmatrix}
\begin{pmatrix}
x^0\\
x^3
\end{pmatrix}, 
\end{eqnarray}
where
\begin{eqnarray}
g^{\alpha\beta} =
\begin{pmatrix}
0 & 2 \\
2 & 0
\end{pmatrix} \,\,\,\,
\text{and} \,\,\,\,\,
g_{\alpha\beta} =
\begin{pmatrix}
0 & \frac{1}{2} \\
\frac{1}{2} & 0
\end{pmatrix}.
\end{eqnarray}
The scalar product can be written as
\begin{equation}
x \cdot p = \frac{1}{2}( x^+ p^- + x^- p^+) - \bm{x}_{\perp} \cdot \bm{p}_{\perp}.
\end{equation}

\subsection{Dirac spinor}

The Dirac spinors in the light-front basis are given by 
\begin{eqnarray}
	u_\lambda(p) &=& \frac{1}{\sqrt{p^+}}(\slashed{p} + m) u_\lambda, \\
	v_\lambda(p) &=& \frac{1}{\sqrt{p^+}}(\slashed{p} - m) v_\lambda,
\end{eqnarray}
where
\begin{eqnarray}
	u_{\uparrow} = \begin{pmatrix}
		1 \\ 0 \\ 0 \\ 0
	\end{pmatrix}, \hspace{1cm} u_{\downarrow}=\begin{pmatrix}
	0 \\ 0 \\ 0 \\ 1
\end{pmatrix}, 
\end{eqnarray}
and $v_\lambda = u_{-\lambda}$.
The antispinors are given by $\bar{u} = u^\dagger \gamma^0$ and $\bar{v} = v^\dagger \gamma^0$.
In the chiral representation, the Dirac matrices are defined as
\begin{eqnarray}
\gamma^0 = \begin{pmatrix} 0  & I \\ I & 0 \end{pmatrix}, \hspace{0.25cm } \gamma^i = \begin{pmatrix} 0  & \sigma^i \\ -\sigma^i & 0 \end{pmatrix}, \hspace{0.25cm}
\gamma^5 = \begin{pmatrix} -I  & 0 \\ 0 & I \end{pmatrix},\quad\quad   
\end{eqnarray}
where $\gamma^\pm = \gamma^0 \pm \gamma^3$ and $\gamma^{R(L)} = \gamma^1 \pm i\gamma^2$. 

\begin{table}[b]
	\centering
 	\begin{ruledtabular}
  		\renewcommand{\arraystretch}{1.8}
	\caption{The Dirac matrix elements $\bar{u}_{\lambda^\prime}(p_1^\prime) \gamma^{\mu} u_{\lambda}(p_1)$ for various components of currents $\mu$, classified by the helicity contribution $\lambda \lambda^\prime$. The helicity flip contributions are zero for plus ($\mu=+$) and transverse [$\mu=R(L)$] components.}
	\label{tab:dirac}
	\begin{tabular}{c c c c c }
        Matrix elements & $\uparrow \uparrow$ & $\uparrow \downarrow$ & $\downarrow \uparrow$ & $\downarrow \downarrow$  \\
        \hline 
        $\bar{u}_{\lambda^\prime}(p_1^\prime) \gamma^+ u_{\lambda}(p_1)$ & $2\sqrt{p_1^+ p_1^{\prime +} } $ & 0 & 0 & $2\sqrt{p_1^+ p_1^{\prime +} } $\\
        $\bar{u}_{\lambda^\prime}(p_1^\prime) \gamma^R u_{\lambda}(p_1)$ & $2p_1^{R}$ & 0 & 0 & $2p_1^{\prime R}$ \\
        $\bar{u}_{\lambda^\prime}(p_1^\prime) \gamma^L u_{\lambda}(p_1)$ & $2p_1^{\prime L}$ & 0 & 0 & $2p_1^{L}$ \\
        \end{tabular}
     \renewcommand{\arraystretch}{1}
   \end{ruledtabular}
\end{table}

Explicitly, the Dirac spinors are
\begin{eqnarray}
	u_{\uparrow}(p) &=& \frac{1}{\sqrt{p^+}}\begin{pmatrix} m \\ 0 \\ p^+ \\ p^R \end{pmatrix}, \quad 
	u_{\downarrow}(p) = \frac{1}{\sqrt{p^+}}\begin{pmatrix} -p^L \\ p^+ \\ 0\\ m \end{pmatrix}, \quad\\
	v_{\uparrow}(p) &=& \frac{1}{\sqrt{p^+}}\begin{pmatrix} -p^L \\ p^+ \\ 0\\ -m \end{pmatrix}, \quad   v_{\downarrow}(p) = \frac{1}{\sqrt{p^+}}\begin{pmatrix} -m \\ 0 \\ p^+ \\ p^R \end{pmatrix}, \quad\quad
\end{eqnarray}
and the antispinors are
\begin{eqnarray}
\bar{u}_{\uparrow}(p) &=& \frac{1}{\sqrt{p^+}}\begin{pmatrix} p^+, & p^L, & m, & 0 \end{pmatrix}, \\
\bar{u}_{\downarrow}(p) &=& \frac{1}{\sqrt{p^+}}\begin{pmatrix} 0,  & m,  & -p^R, & p^+ \end{pmatrix}, \\
\bar{v}_{\uparrow}(p) &=& \frac{1}{\sqrt{p^+}}\begin{pmatrix} 0,  & -m,  & -p^R, & p^+  \end{pmatrix},  \\
\bar{v}_{\downarrow}(p) &=& \frac{1}{\sqrt{p^+}}\begin{pmatrix} p^+, & p^L, & -m, & 0  \end{pmatrix}.	
\end{eqnarray}
The elements of Dirac matrix can be derived and are summarized in Table~\ref{tab:dirac}.

\subsection{Spin-orbit wave function}

The explicit form of the spin-orbit wave functions for the pseudoscalar and vector mesons are given by
\begin{equation}
\mathcal{R}^{00}_{\lambda_1\lambda_2} (x,\bm{k}_\perp) 
=\frac{\mathcal{R}_0}{\sqrt{2}} 
\begin{pmatrix}
-k^L 		& \mathcal{A}\\
-\mathcal{A} & -k^R\\
\end{pmatrix},
\end{equation}
and
\begin{eqnarray}
\mathcal{R}^{1+1}_{\lambda_1 \lambda_2} (x,\bm{k}_\perp)  &=& \mathcal{R}_0
\begin{pmatrix}
\mathcal{A} + \frac{k_\bot^2}{\mathcal{D}_0} & k^R \frac{\mathcal{M}_1}{\mathcal{D}_0}\\
- k^R \frac{\mathcal{M}_2}{\mathcal{D}_0}    & -\frac{(k^R)^2}{\mathcal{D}_0} 	\\
\end{pmatrix},\\
\mathcal{R}^{10}_{\lambda_1 \lambda_2} (x,\bm{k}_\perp)  &=& \frac{\mathcal{R}_0}{\sqrt{2}}
\begin{pmatrix}
k^L \frac{\mathcal{M}}{\mathcal{D}_0} & \mathcal{A} + \frac{2k_\bot^2}{\mathcal{D}_0} \\
\mathcal{A} + \frac{2k_\bot^2}{\mathcal{D}_0}   &   -k^R \frac{\mathcal{M}}{\mathcal{D}_0} 	\\
\end{pmatrix},\quad \\
\mathcal{R}^{1-1}_{\lambda_1 \lambda_2} (x,\bm{k}_\perp)  &=& \mathcal{R}_0
\begin{pmatrix}
-\frac{(k^L)^2}{\mathcal{D}_0} & k^L \frac{\mathcal{M}_2}{\mathcal{D}_0}\\
- k^L \frac{\mathcal{M}_1}{\mathcal{D}_0}    &   \mathcal{A} + \frac{k_\bot^2}{\mathcal{D}_0} 	\\
\end{pmatrix},
\end{eqnarray}
with $k^{R(L)}=k_x\pm i k_y$ and
$\mathcal{R}_0 =1/\sqrt{\mathcal{A}^2 + \bm{k}_\perp^2}$. 
Furthermore, we define $\mathcal{A}=xm_{\bar{q}}+(1-x)m_q$ and $ \mathcal{M}=\mathcal{M}_2-\mathcal{M}_1$ with
\begin{eqnarray}
    \mathcal{M}_1 &=& x M_0 + m_1, \\
    \mathcal{M}_2 &=& (1 - x) M_0 + m_2,
\end{eqnarray}
Note that when computing the transition operator, we need to take the Hermitian conjugate, ensuring that $(k^{R(L)})^\dagger = k^{L(R)}$.

\subsection{Antisymmetric tensor}

The antisymmetric Levi-Civita tensor is used to compute the Lorentz structure in Eq.~\eqref{eq:rad_M1}.
It is defined as
\begin{eqnarray}
\epsilon^{\mu\nu\rho\sigma} = \frac{1}{\sqrt{-{\rm det} (g)}} 
\begin{cases}
+1 & {\rm even\ permutation\ } -+xy, \\
-1 & {\rm odd\ permutation\ } -+xy, \\
 0 & {\rm others,\ } \\
\end{cases} \nonumber\\
\end{eqnarray}
where $\sqrt{-{\rm det} (g)}=1/2$. 

\section{Matrix elements of radiative decays}
\label{app:radiative}

We provide a detailed calculation of the matrix elements for the radiative decays in the LFQM, 
considering various combinations of current components and polarizations.

\subsection{Plus component $(\mu=+)$ and transverse polarization $(h=\pm1)$ }

The matrix elements for plus component and transverse polarization can be written as
\begin{equation}
\mathcal{S}^{+}_h = \frac{1}{x} \sum_{\lambda \lambda^\prime \bar{\lambda}} \mathcal{R}_{\lambda^\prime \bar{\lambda}}^{00\dagger}(x,\bm{k}_\perp^\prime) \bar{u}_{\lambda^\prime}(p_1^\prime) \gamma^+ u_{\lambda}(p_1) \mathcal{R}_{\lambda \bar{\lambda}}^{11} (x,\bm{k}_\perp).
\end{equation}
Because the helicity flip terms in the Dirac matrix elements vanish, namely, 
\begin{eqnarray}
    \bar{u}_{\lambda^\prime}(p_1^\prime) \gamma^+ u_{\lambda}(p_1) = 2\sqrt{p_1^+p_1^{\prime +}}\delta_{\lambda\lambda^\prime},
\end{eqnarray}
the summation is reduced to a sum over $\lambda$ and $\bar{\lambda}$. 
The remaining terms, defined by $\mathcal{S}_{+1}^{+}(\lambda\bar{\lambda})$, are given by
\begin{eqnarray}
\mathcal{S}_{+1}^{+}(\uparrow\uparrow) &=& -\sqrt{2}P^+N_0 k^{\prime R} \left(\mathcal{A} + \frac{k_\bot^2}{\mathcal{D}_0}\right),  \\
\mathcal{S}_{+1}^{+}(\downarrow\downarrow) &=& \sqrt{2}P^+N_0 k^{\prime L}\frac{k^R k^R}{\mathcal{D}_0}, \\
\mathcal{S}_{+1}^{+}(\uparrow\downarrow) &=& \sqrt{2}P^+N_0 \mathcal{A}k^R \frac{\mathcal{M}_1}{\mathcal{D}_0}, \\
\mathcal{S}_{+1}^{+}(\downarrow\uparrow) &=& \sqrt{2}P^+N_0 \mathcal{A}k^R \frac{\mathcal{M}_2}{\mathcal{D}_0} , 
\end{eqnarray}
with 
\begin{eqnarray}
    N_0 = \frac{1}{\sqrt{\mathcal{A}^2+\bm{k}_\perp^2} \sqrt{\mathcal{A}^2+\bm{k}_\perp^{\prime 2}} }.
\end{eqnarray}
By summing all four terms above, we obtain
\begin{eqnarray}
\frac{\mathcal{S}^{+}_{+1}}{N_0} &=& \sqrt{2}P^+(1-x) \left[ \mathcal{A}q^R +\dfrac{\left( q^R k^L k^R -q^L k^R k^R\right)}{\mathcal{D}_0}  \right]. \nonumber \\
\end{eqnarray}
Note that the $\mathcal{S}^{+}_{-1}$ term can be computed similarly.
We then take the average of the sum of matrix elements as $\mathcal{S}^+_{\pm1} = \frac{1}{2}(\mathcal{S}^{+}_{+1} q^L + \mathcal{S}^{+}_{-1} q^R)$.
By combining the tensor and matrix transition terms, we obtain the operator 
\begin{eqnarray}
    \mathcal{O}_{\pm1}^+ &=& \frac{\sqrt{2}}{\bm{q}^2_\perp P^+}\frac{\mathcal{S}^+_{\pm1}}{N_0}\nonumber\\
    &=& 2(1-x)\left[ \mathcal{A}  +\frac{2}{\mathcal{D}_0} \left( \bm{k}_\perp^2 -\frac{(\bm{k}_\perp \cdot \bm{q}_\perp)^2 }{\bm{q}_\perp^2} \right) \right] ,\quad \quad 
\end{eqnarray}
where we have used $q^Lq^R=\bm{q}_\perp^2$ and 
\begin{eqnarray}
    \frac{1}{2}[k^L q^R + k^R q^L ] &=& \bm{k}_\perp \cdot \bm{q}_\perp,\\
    \frac{1}{2}[k^L k^L q^R q^R + k^R k^R q^L q^L ] &=& 2(\bm{k}_\perp \cdot \bm{q}_\perp)^2 -  \bm{k}_\perp^2 \bm{q}_\perp^2.\nonumber\\ 
\end{eqnarray}

\subsection{Transverse component $[\mu=R(L)]$ and longitudinal polarization $(h=0)$}

The matrix elements for transverse component and longitudinal polarization can be written as
\begin{equation}
\mathcal{S}^{R(L)}_0 = \frac{1}{x}\sum_{\lambda^\prime \lambda \bar{\lambda}} \mathcal{R}_{\lambda^\prime \bar{\lambda}}^{00\dagger}(x,\bm{k}_\perp^\prime) \bar{u}_{\lambda^\prime}(p_1^\prime) \gamma^{R(L)} u_{\lambda}(p_1)
\mathcal{R}_{\lambda \bar{\lambda}}^{10} (x,\bm{k}_\perp),
\end{equation}
where we introduce the indices $R(L)$ for the transverse components. 
For $\mu=R$, the remaining terms, defined by $\mathcal{S}_{0}^{R}(\lambda\bar{\lambda})$, are given by
\begin{eqnarray}
\mathcal{S}^{R}_0(\uparrow\uparrow) &=& -\frac{N_0}{x} p_1^R k^{\prime R} \frac{\mathcal{M}k^L}{\mathcal{D}_0}, \\
\mathcal{S}^{R}_0(\downarrow\downarrow) &=&\frac{N_0}{x} p_1^{\prime R} k^{\prime L}\frac{\mathcal{M}k^R}{\mathcal{D}_0}, \\
\mathcal{S}^{R}_0(\uparrow\downarrow) &=& \frac{N_0}{x} p_1^R \mathcal{A}\left(\mathcal{A}+\frac{2\bm{k}^2}{\mathcal{D}_0}\right), \\
\mathcal{S}^{R}_0(\downarrow\uparrow) &=& -\frac{N_0}{x} p_1^{\prime R}\mathcal{A}\left(\mathcal{A}+\frac{2\bm{k}^2_{\perp}}{\mathcal{D}_0}\right).
\end{eqnarray}
Summing them up, we find
\begin{eqnarray}
\frac{\mathcal{S}^{R}_0}{N_0} &=& \frac{1}{x} \left[ \frac{\mathcal{M}}{\mathcal{D}_0} (-k^R k^{\prime R} k^L + k^R k^{\prime L} k^R - q^Rk^{\prime L} k^R  ) \right. \nonumber \\
&&\left. +\mathcal{A}\left(\mathcal{A}+\frac{2k_\perp^2}{\mathcal{D}_0}\right) q^R  \right].
\end{eqnarray}
The $\mathcal{S}^{L}_0$ can be computed similarly.
Taking the average of those two components, we have $\mathcal{S}_0^{\perp} = \frac{1}{2} (q^L \mathcal{S}^{R}_0 - q^R \mathcal{S}^{L}_0).$
By combining the tensor and matrix transition terms,  
we derive the operator 
\begin{eqnarray}
    \mathcal{O}_{0}^\perp &=& \frac{1}{\bm{q}^2_\perp M_0}\frac{\mathcal{S}^\perp_{0}}{N_0}\nonumber\\
    &=& \frac{1}{xM_0}\left\{ \mathcal{A}\left(\mathcal{A}+\frac{2\bm{k}_\perp^2}{\mathcal{D}_0}\right) + \frac{\mathcal{M}}{\mathcal{D}_0} \biggl[ (1-2x) \bm{k}_\perp^2 \right. \nonumber \\
&& \left. +(1-x) \left( (\bm{k}_\perp\cdot \bm{q}_\perp) -\frac{2(\bm{k}_\perp\cdot \bm{q}_\perp)^2}{\bm{q}_\perp^2}  \right)\biggr]\right\}. \quad\quad\quad 
\end{eqnarray}

\subsection{Transverse component $[\mu=R(L)]$ and transverse polarization $[h=+1(-1)]$}

Next, we compute the matrix elements for transverse component and polarization, which are given by
\begin{equation}
\mathcal{S}^{R(L)}_{\pm1} = \frac{1}{x}\sum_{\lambda^\prime \lambda \bar{\lambda}} \mathcal{R}_{\lambda^\prime \bar{\lambda}}^{00\dagger}(x,\bm{k}_\perp^\prime) \bar{u}_{\lambda^\prime}(p_1^\prime) \gamma^{R(L)} u_{\lambda}(p_1)
\mathcal{R}_{\lambda \bar{\lambda}}^{1\pm1} (x,\bm{k}_\perp).
\end{equation}
When computing the matrix elements, we realize that the results depend on whether the assignment $\mu = R$ is combined with $h = +1$ or $h = -1$. 
This dependence is also reflected in the computation of the Lorentz structure, as shown in Table~\ref{tab:tensor}.
Here we compute the first combination of $\mu=R(L)$ and $h=+1(-1)$.
For $\mu =R$ and $h=+1$, the surviving terms, defined as $\mathcal{S}_{+1}^{R}(\lambda\bar{\lambda})$, are given by
\begin{eqnarray}
\mathcal{S}_{+1}^{R}(\uparrow\uparrow) &=& -\sqrt{2} N_0 p_1^{R}k^{\prime R}  \left( \mathcal{A} +  \frac{\bm{k}_{\perp}^2}{\mathcal{D}_0} \right),  \\
\mathcal{S}_{+1}^{R}(\downarrow\downarrow) &=& \sqrt{2} N_0 p_1^{\prime R} k^{\prime L} \frac{k^R k^R}{\mathcal{D}_0}, \\
\mathcal{S}_{+1}^{R}(\uparrow\downarrow) &=& \sqrt{2} N_0 \mathcal{A} p_1^{R} k^{R} \frac{\mathcal{M}_1}{\mathcal{D}_0}, \\
\mathcal{S}_{+1}^{R}(\downarrow\uparrow) &=& \sqrt{2} N_0 \mathcal{A} p_1^{\prime R} k^R \frac{\mathcal{M}_2}{\mathcal{D}_0}.
\end{eqnarray}
Summing them up, we obtain
\begin{eqnarray}
\frac{\mathcal{S}^{R}_{+1}}{N_0} &=& \sqrt{2}\left[ 
\frac{\mathcal{A}}{\mathcal{D}_0} \left\{k^R k^R (\mathcal{M}_1 + \mathcal{M}_2 ) \right. \right.  \nonumber \\
&&\left. \left. - q^R k^R \mathcal{M}_2 + xP^R(\mathcal{M}_1 + \mathcal{M}_2 ) \right\} \right. \nonumber \\
&&\left. - (k^R + xP^R)(k^R - (1-x)q^R) \left( \mathcal{A} +  \frac{\bm{k}_{\perp}^2}{\mathcal{D}_0} \right) \right. \nonumber \\
&&\left. + k^R k^{\prime L} \frac{k^R k^R}{\mathcal{D}_0} - q^R k^{\prime L} \frac{k^R k^R}{\mathcal{D}_0} + xP^R k^{\prime L} \frac{k^R k^R}{\mathcal{D}_0} \right], \nonumber \\
\end{eqnarray} 
and $\mathcal{S}^{L}_{-1}$ can be computed similarly.
We then combine them to obtain $\mathcal{S}_{\pm1}^{\perp} = \frac{1}{2} (P^{L}q^{L}\mathcal{S}^{R}_{+1} + P^{R}q^{R}\mathcal{S}^{L}_{-1})$.
The operator can be computed as
\begin{eqnarray} \label{eq:RL1}
\mathcal{O}_{\pm1}^\perp &=& -\frac{\sqrt{2}}{\bm{P}_\perp^2 \bm{q}_\perp^2}\mathcal{S}_{\pm1}^{\perp}, \nonumber\\
&=& 2(1-x)\left[ \mathcal{A}  +\frac{2}{\mathcal{D}_0} \left( \bm{k}_\perp^2 -\frac{(\bm{k}_\perp \cdot \bm{q}_\perp)^2 }{\bm{q}_\perp^2} \right) \right] \nonumber\\
&& + \tilde{\mathcal{O}}(\bm{P}_\perp), 
\end{eqnarray}
where we set $P_\perp \to 0$ to drop the $\bm{P}_\perp$ dependent part $\tilde{\mathcal{O}}(\bm{P}_\perp)$ for the sake of simplicity.
It is evident that the result is the same as that with the plus current $(\mu=+)$.

\subsection{Transverse component $[\mu=R(L)]$ and transverse polarization $[h=-1(+1)]$}

Lastly, we compute the matrix elements with the second combinations of $\mu=R(L)$ and $h=-1(+1)$.
For $\mu = L$ and $h=+1$, the remaining terms are given by
\begin{eqnarray}
    \mathcal{S}_{+1}^{L}(\uparrow\uparrow) &=& -\sqrt{2} N_0 p_1^{L} k^{\prime R} \left( \mathcal{A} +  \frac{\bm{k}_{\perp}^2}{\mathcal{D}_0} \right), \\
    \mathcal{S}_{+1}^{L}(\downarrow\downarrow) &=& \sqrt{2} N_0 p_1^{L} k^{\prime L} \frac{k^R k^R}{\mathcal{D}_0}, \\
    \mathcal{S}_{+1}^{L}(\uparrow\downarrow) &=& \sqrt{2} N_0 \mathcal{A} p_1^{\prime L} k^{R} \frac{\mathcal{M}_1}{\mathcal{D}_0}, \\
    \mathcal{S}_{+1}^{L}(\downarrow\uparrow) &=&  \sqrt{2} N_0 \mathcal{A} p_1^{L} k^R \frac{\mathcal{M}_2}{\mathcal{D}_0}.
\end{eqnarray}
Summing all four terms in $\mathcal{S}^{L}_{+1}$, we obtain
\begin{eqnarray}
     \frac{\mathcal{S}^{L}_{+1}}{N_0} &=& \sqrt{2} \left[ k^R q^L \left(\mathcal{A} + \frac{xk_{\perp}^2}{\mathcal{D}_0} - \frac{\mathcal{A}\mathcal{M}_1}{\mathcal{D}_0} \right) \right. \nonumber \\
     && \left. + (1-x) ( k^L q^R - q_{\perp}^2 )\left(\mathcal{A} + \frac{k_{\perp}^2}{\mathcal{D}_0}\right) \right]\quad \quad 
\end{eqnarray}
and $\mathcal{S}^{R}_{-1}$ can be computed similarly. 
We combine them to obtain $\mathcal{S}_{\pm1}^{\perp} = \frac{1}{2} (\mathcal{S}^{R}_{-1} + \mathcal{S}^{L}_{+1})$. 
We then find the operator as
\begin{eqnarray}
    \mathcal{O}^{\perp}_{\pm 1} &=& \frac{\sqrt{2}}{q^- P^+}\mathcal{S}_{\pm1}^{\perp}, \nonumber\\
    &=& \dfrac{2}{x(M_0^2-M_0^{\prime 2} -\bm{q}_\perp^2)} \nonumber\\
    && \left[ (\bm{k}_\perp\cdot\bm{q}_\perp) \left(\mathcal{A} + \dfrac{x \bm{k}_{\perp}^2}{\mathcal{D}_0} - \dfrac{\mathcal{A}\mathcal{M}_1}{\mathcal{D}_0} \right) \right.\nonumber\\
    &&\left. + (1-x)(\bm{k}_\perp\cdot\bm{q}_\perp -\bm{q}_\perp^2) \left(\mathcal{A} + \dfrac{\bm{k}_{\perp}^2}{\mathcal{D}_0}\right) \right].\quad \quad 
\end{eqnarray}
Although they are derived from the transverse component and polarization, 
we show that the above formula is different from that in Eq.~\eqref{eq:RL1}.

\bibliography{reference}

\end{document}